\documentclass{aa}
\usepackage[margin=2cm]{geometry}                % See geometry.pdf to learn the layout options. There are lots.
\usepackage{graphicx}
\usepackage{amsmath,amssymb}
\usepackage{xcolor}
\usepackage{epstopdf}
\usepackage{natbib}
\usepackage{hyperref}
\hypersetup{
    colorlinks=true,
    linkcolor=blue,
    citecolor=blue,
    urlcolor=cyan,
    }
\usepackage{txfonts}
\usepackage{url}

\newcommand{\myvect}[1]{\mathbf#1}
\newcommand{\myhat}[1]{\mathbf#1}

\begin{document}

\title{Forecasting the solar cycle using variational data assimilation: validation on cycles 22 to 25}

\author{L. Jouve\inst{\ref{irap1}},
C.P. Hung\inst{\ref{cea},\ref{ipgp}}, A. S. Brun\inst{\ref{cea}}, S. Hazra\inst{\ref{cea}}, A. Fournier\inst{\ref{ipgp}}, O. Talagrand\inst{\ref{lmd}}, B. Perri\inst{\ref{cea}}, A. Strugarek\inst{\ref{cea}} }

\institute{Institut de Recherche en Astrophysique et Plan\'etologie, Universit\'e de Toulouse, CNRS, CNES, UPS, 14 avenue Edouard Belin 31400 Toulouse, France\\
email:laurene.jouve@irap.omp.eu\label{irap1}
\and D\'epartement d'Astrophysique/AIM 
CEA/IRFU, CNRS/INSU, Univ. Paris-Saclay \& Univ. de Paris Cit\'e 
91191 Gif-sur-Yvette, France \label{cea}
\and Universit\'e Paris Cit\'e, Institut de physique du globe de Paris, CNRS, F-75005 Paris, France \label{ipgp}
\and Laboratoire de m\'et\'eorologie dynamique, UMR 8539, Ecole Normale Sup\'erieure, Paris Cedex 05, France\label{lmd} }

   \date{Received; accepted }

\abstract{Forecasting future solar activity has become crucial in our modern world, where intense eruptive phenomena mostly occurring during solar maximum are likely to be strongly damaging to satellites and telecommunications. It is however a very difficult task owing to the highly turbulent flows existing in the solar interior.}{We present a 4D variational assimilation technique applied for the first time to real solar data, consisting in the time series of the sunspot number and the line-of-sight surface magnetic field from 1975 to 2024.  Our method is tested against observations of past cycles 22, 23, 24 and on the ongoing cycle 25 for which we give an estimate of the imminent maximum value and timing and also provide a first forecast of the next solar minimum.}{We use a variational data assimilation technique applied to a solar mean-field Babcock-Leighton flux-transport dynamo model. This translates into the minimization of an objective function with respect to the control vector defined here by a set of coefficients representing the meridional flow and the initial magnetic field. Ensemble predictions are produced in order to obtain uncertainties on the timing and value of the maximum of cycle $n+1$, when data on cycle $n$ is assimilated. We study in particular the influence of the phase during which data is assimilated in the model and of the weighting of various terms in the objective function.}{The method is validated on cycles 22, 23 and 24 with very satisfactory results. We find in particular good convergence of our predictions (both in accuracy and precision) when the assimilation window encompasses more and more of the rising phase of cycle $n+1$. For cycle 25, predictions vary again depending on the extent of the assimilation window but start converging past 2022 to a solar maximum reached between mid-2024 up to the beginning of 2025 with a sunspot number value of $143.1 \pm 15.0$. Relatively close values of the maximum are found in both hemispheres within a time lag of a few months. We also forecast a next minimum around late 2029, with still significant errorbars.}{The data assimilation technique presented here combining a physics-based model and real solar observations produces promising results for future solar activity forecasting.} 

\titlerunning{4D-Var applied to the solar cycle}
\authorrunning{L. Jouve et al.}
      
\maketitle

\section{Introduction}
The multitude of manifestations of solar activity have important consequences on the Sun's environment including our planet and the thousands of active satellites orbiting around it. These manifestations include a changing solar flux irradiating the Earth, solar energetic particles, coronal mass ejections (CMEs) and flares, magnetic fields and solar wind variabilities on various timescales. In the connected world we now live in, solar activity and the potential consequences on satellites and telecommunications infrastructures have become crucial to understand and predict. Whether one is interested in space weather (short time-scales events as CMEs) or space climate (longer term evolution of the Sun's cyclic activity), the question always comes back to the understanding of the production of solar magnetic fields which govern the Sun's activity and heliosphere. 

The Sun is thought to operate a non-linear turbulent dynamo producing magnetic fields with a huge range of temporal and spatial scales, making it extremely difficult to model and predict in all its details. However, if we are interested in the forecasting of the characteristics of the 11-year sunspot cycle, we may focus on the large-scale features mostly, such as the mainly dipolar field exhibited at solar minimum peaking at the solar poles and the sunspots appearing periodically at the solar surface and thought to be the product of toroidal fields emerging from the solar interior. 

The predictions of the sunspot number (SSN) evolution has been an active field of research for several decades now. Various approaches have been adopted with methods which can be classified into three main groups: precursor methods, extrapolation methods and physics-based methods \citep[see][for details on those methods and results]{Petrovay2020}. It is still unclear if some of these methods are superior to others but the relatively recently-developed physics based methods (relying on dynamo and/or flux-transport models: e.g. \citet{Dikpati06, Dikpati12, Dikpati14, Choudhuri07, Bhowmik2018, Labonville2019}) have the advantage to include physical processes thought to play a key role in the solar interior and to enable the assimilation of observed data \citep[see][]{Nandy2021}. In this work, we choose to use one of those physics-based models in which data is assimilated through a 4D-var technique routinely used today to predict weather on Earth \citep{Klinker2000}. 

Following previous papers \citep{Hungetal2015ApJ, Hungetal2017ApJ}, we apply a 4D variational data assimilation (DA) technique to a mean-field flux-transport Babcock-Leighton (BL) model, which has been used for decades to reproduce the evolution of the large-scale features of the solar cycle \citep{Charbonneau2010lrsp}. In those previous works, the BL model produced synthetic observations on which various levels of Gaussian noise were added. Our 4D-Var assimilation technique, which necessitates the development of an adjoint model \citep{Talagrand87, Fournier10}, enabled our physical model to recover to a very good accuracy the meridional flow and initial magnetic conditions which were used to produce the synthetic data.

In this work, we move one step forward by applying our technique to real solar data, consisting in the time series of SSN and surface line-of-sight magnetic fields provided by different observatories. Another addition is the possibility to predict not only the amplitude and timing of the next maximum but also to have some insight on the asymmetry between the Northern and Southern hemispheres which sometimes reach their peak SSN at very different times \citep[see e.g.][for a focus on North/South asymmetries in the solar cycle]{Waldmeier71, Berdyugina03, Temmer06, Norton10,Finley2023}. The goal of our work is thus now to minimize an objective function measuring the misfit between outputs of our model and associated real solar observations. To minimize the objective function, a control vector is adjusted by the assimilation step of our method. Similarly to \citet{Hungetal2017ApJ}, the control vector consists of coefficients defining the meridional circulation and initial conditions of the magnetic field. Our method, which will be referred in the paper as \emph{the solar predict tool} is applied to the 3 previous solar cycles (namely cycles 22, 23 and 24) and to the ongoing cycle 25 whose peak amplitude is, as of writing, very close to being reached. We show a very satisfactory agreement for all the previous 3 cycles and then estimate both the current sunspot maximum and the timing of the next minimum of solar activity.

The paper is organized as follows: the methodology (namely the model and data) is first presented in Section \ref{sec_method}. Section \ref{sec_validation} then shows the tests performed to validate our method on the 3 previous cycles. The main results for the forecasting of cycle 25 are then presented in Section \ref{sec_cycle25} and we finally conclude in Section \ref{sec_conclu}.

\section{Methodology}
\label{sec_method}
\subsection{Dynamo model}
\label{sec_model}

Similarly to \cite{Hungetal2017ApJ}, we base our 11-yr solar cycle forecasts on a mean field Babcock-Leighton flux transport dynamo model. We here provide the main elements of this 2.5D magneto-hydrodynamical (MHD) model. Expanding the assumed axisymmetric magnetic field as:
$$
{\bf B}=B_\phi {\bf e_\phi} + {\bf \nabla} \times (A_\phi {\bf e_\phi})
$$

\noindent we solve for the following mean field dynamo equations for the poloidal potential $A_{\phi}$ and the toroidal magnetic field $B_{\phi}$: 

\begin{equation}\label{eq:Adyn}
\begin{split}
\partial_t A_{\phi} & =\frac{\eta}{\eta_t}  \left(\nabla^2 - \frac{1}{\varpi^2}\right) A_{\phi}
-Re \frac{\myvect{v}_p}{\varpi}\cdot \nabla (\varpi A_{\phi}) % \\
  + C_sS(r, \theta, B_{\phi}), 
\end{split}
\end{equation}

\begin{equation}\label{eq:Bdyn}
\begin{split}
\partial_t B_{\phi} = &
\frac{\eta}{\eta_t}  \left(\nabla^2 - \frac{1}{\varpi^2}\right) B_{\phi}
+\frac{1}{\varpi} \frac{\partial (\varpi B_{\phi})}{\partial r} \frac{\partial (\eta/\eta_t)}{\partial r} \\ %\\
&-Re\, \varpi \myvect{v}_p\cdot \nabla \left( \frac{B_{\phi}}{\varpi }\right)
- Re\, B_{\phi} \nabla \cdot \myvect{v}_p \\
& + C_{\Omega} \varpi \left[ \nabla \times (A_{\phi} \myhat{e_{\phi}}) \right] \cdot \nabla \Omega ,
\end{split}
\end{equation}
where $\varpi=r \sin \theta$, $\myvect{v}_p$ is the meridional circulation, ${\bf \Omega}=\Omega\, {\bf e}_z$ is the angular velocity, $\eta$ is the magnetic diffusivity
and $S$ is the source of the poloidal field at the solar surface (note that we do not include an $\alpha$-effect). 

The length is normalized with the solar radius $R_{\sun}$, time is normalized using the diffusive time scale $R_{\sun}^2/{\eta}_t$ 
where ${\eta}_t$ is the envelope turbulent diffusivity. 
We then introduce three dimensionless parameters, namely the Reynolds number based on the meridional flow speed $Re=v_o R_{\sun}/\eta_t$, 
the strength of the Babcock-Leighton source $C_s=s_o R_{\sun}/\eta_t$ 
and the strength of the $\Omega$-effect $C_{\Omega}=\Omega_oR_{\sun}^2/\eta_t$. For the reference model used in this work, the following values are adopted for these dimensionless parameters: $Re=100$, $C_s=35$ and $C_{\Omega}=9.27 \times 10^3$, translating into dimensioned values of the equatorial rotation rate of $\Omega_o = 2\pi \times 456 \, \mathrm{nHz}$, a maximal surface flow speed of $v_o=21.5 \, \rm m.s^{-1}$, an amplitude of the BL source $s_o=7.54 \, \rm m.s^{-1}$ and a turbulent diffusivity at the surface $\eta_t=1.5\times 10^{12} \, \rm cm^2.s^{-1}$.

We use the same model ingredients as we did in \cite{Hungetal2017ApJ}, namely a 2-step profile in the radial direction for the resistivity profile:
\begin{eqnarray}
\label{eq:etaprofile}
\frac{\eta}{\eta_t}=\frac{\eta_c}{\eta_t}&+&\frac{\eta_m-\eta_t}{2\eta_t}\left[1 +\tanh\left(\frac{r-r_{bcz}}{d_1}\right)\right]\\ \nonumber
&+&\frac{\eta_t-\eta_m}{2\eta_t}\left[1+\tanh\left(\frac{r-r_2}{d_1}\right)\right],
\end{eqnarray}
where $\eta_c=10^{12}$ cm$^2$~s$^{-1}$, $\eta_m=10^{11}$ cm$^2$~s$^{-1}$, $\eta_t=1.5\times 10^{12}$ cm$^2$~s$^{-1}$, $r_{bcz}=0.72$ (representing the base of the convection zone in our model), $r_2=0.95$, $d_1=0.016$.
Using this 2-step resistivity profile with a high diffusion at the surface helps lowering 
the ratio of radial magnetic field at the pole to that near the equator \citep{HottaYokoyama2010}. A high value of the diffusivity at the bottom helps to produce a more realistic profile for the magnetic cycle, in particular the tendency of strong cycles to rise faster than weak ones, the so-called Waldmeier-effect \citep{Karak2011}. We however do not anticipate a strong impact of this particular choice of a high diffusivity on the results of the assimilation procedure.s

The expression of the non-local Babcock-Leighton source term reads:
\begin{eqnarray}
\label{eq:sprofile}
S(r,\theta,B_\phi)&=&\frac{1}{2}\left[1+\tanh \left(\frac{r-r_2}{d_2}\right)\right] \left[1-\tanh \left (\frac{r-1}{d_2}\right)\right]\\ \nonumber
& \times & \cos\theta \sin\theta \left[1+\left(\frac{B_\phi(r_{bcz},\theta,t)}{B_0}\right)^2 \right]^{-1}\\ \nonumber
&\times & B_\phi(r_{bcz},\theta,t)
\end{eqnarray}
\noindent  where $B_\phi(r_{bcz},\theta,t)$ represents the toroidal field at the base of the convection zone, the thickness $d_2=0.008$ and the radii $r_2$ and $r_{bcz}$ are defined above. We here choose a low value of the quenching parameter $B_0=9\,\rm G$ in order to have surface radial field values directly comparable to the observations (see Fig.\ref{fig:obsbtfy}). Indeed, a more conventional value of $B_0=10^4 \, \rm G$ always produces unrealistically strong values of the surface radial field at the poles, as reported in a number of previous studies \citep[see for example the review of][]{Charbonneau2010lrsp}. 

We now turn to the meridional circulation profile. 
We express the flow in the convection zone as the curl of the stream function $\psi$:
 \begin{equation}\label{eq:curlvp}
  \myvect{v}_p(r, \theta,t)= \nabla \times (\psi(r,\theta,t) \myhat{e_{\phi}}).
 \end{equation}

\noindent We specify the explicit expression of $\psi(r,\theta, t)$ in this case in terms of its expansion on 
a chosen set of basis functions, namely Legendre polynomials in latitude and sine functions in radius:
 \begin{equation}\label{eq:MC}
 \begin{split}
 %\MoveEqLeft
 & \psi (r,\theta,t)= -\frac{2}{\pi}\left(\frac{r-r_{mc}}{1-r_{mc}}\right)^{2.5}(1-r_{mc}) \\
 & \times
   \begin{cases}
      \sum\limits_{k=1}^{m} \sum\limits_{l=1}^{n} d_{k,l}(t) \sin \left[ \frac{k\pi (r-r_{mc})}{1-r_{mc}} \right]
 %                                \\ \times
  P_l^1 (-\cos \theta) & \text{if }r_{mc} \leq r \leq 1 \\
      0 & \text{if } 0.6 \leq r<r_{mc},
   \end{cases}
 \end{split}
 \end{equation}
where $P^1_l$ is the associated Legendre polynomials of degree $l$ and order 1. 
The meridional flow is allowed to penetrate to a radius $r_{mc}=0.65$, 
i.e. slightly below the base of the convection zone.
This profile is used to produce a 22-year magnetic cycle dynamo model with a surface flow $\sim 20$ ms$^{-1}$ \citep{Yeates2008}, 
consistent with helioseismic observations \citep{Ulrich2010, BasuMC23cycle10, Komm2015SolarPhys}. 
We note here that the expansion coefficients $d_{k,l}(t)$ are obtained via our data assimilation procedure explained in the next section and are modulated in time so that the meridional flow is time-dependent.

Finally, the numerical domain is $(r,\theta) \in [0.6,1] \times [0,\pi]$. The grid size is $n_r \times n_{th} = 129 \times 129$, and the time step is $10^{-6}$, equivalent to $0.112$ day. 
The toroidal field $B_{\phi}=0$ at the boundary of the domain, and for $A_{\phi}$, 
we impose the pure radial field approximation at the surface, i.e., $\partial_r (r A_{\phi})=0$ at $r=1$, 
and $A_{\phi}=0$ at all the other boundaries. We note that we used radial boundary conditions for simplicity in this work but the impact of more realistic conditions (as a match to an external potential field) would be worth considering in the future.

\subsection{Solar data and pre-processing}

As explained in the next section, our data assimilation  procedure uses two types of solar data sets: a) the sunspot number time series and b) the line of sight surface magnetic field ($B_{los}$) from monthly generated synoptic maps. 

More precisely, for the SSN, we use the new (v2.0) time series provided by the \href{https://www.sidc.be/SILSO/home}{SIDC SILSO data base}. Since, as explained later, the \emph{solar predict tool} can assimilate hemispherical data, we use in practice the hemispherical data base from SILSO. The data base runs from 1992 until present. In order to be able to cover the same time period than $B_{los}$, we extended in the past the SILSO hemispherical data series with the UCLE station operated at the Royal Observatory of Brussels. This allows to provide a hemispherical SSN time series from 1975 until present (see Figure \ref{fig:obsssn}). We show in the appendix how we cross-calibrated the two times series around the 1992 period. This data is used in the objective function  $\mathcal{J}_{W}$ (that will be described later) through the variables $W_{N}^o$ and $W_{S}^o$. We use the 13-month smoothed SSN time series for the assimilation when minimizing the objective function $\mathcal{J}_{W}$. We also use the monthly average time series for plotting purposes but its higher monthly temporal variation is too irregular to be used efficiently in the data assimilation procedure. We note that using the 13-month smoothed SSN data series is standard in the community. Indeed, this SSN series is the reference used to officially determine the amplitudes of the minima and maxima and the timing of each solar cycle. Further note that this results in having a 6-month time lag when assimilating the data for forecasting cycle 25, as this cycle is still ongoing. At the time of writing of this paper (Fall 2024) this means that we assimilate SSN data up to April 2024.

\begin{figure}[h!]
\begin{center}
\includegraphics[width=0.48\textwidth]{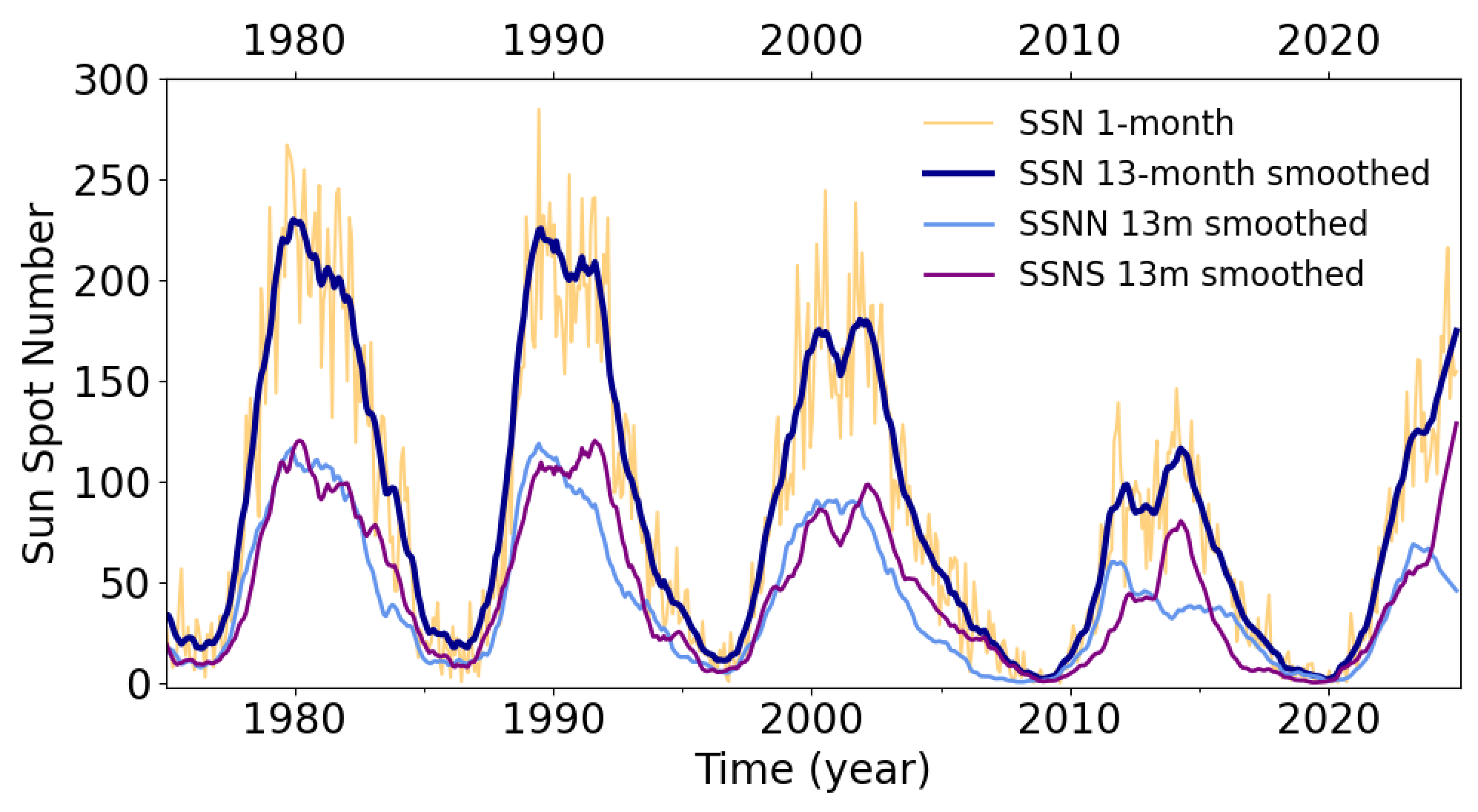}
\end{center}
\caption{SSN time series used in the assimilation procedure. We assimilate the hemispherical data and can choose to either forecast the total or hemispherical solar activity. Shown in dark blue is the total 13-month smoothed SSN times series and in light blue shades the northern and southern sub time series (north + south = total). We superimposed in orange the monthly data, which is 6 months ahead by construction.}
\label{fig:obsssn}
\end{figure}

\begin{figure}[h!]
\begin{center}
\includegraphics[width=0.48\textwidth]{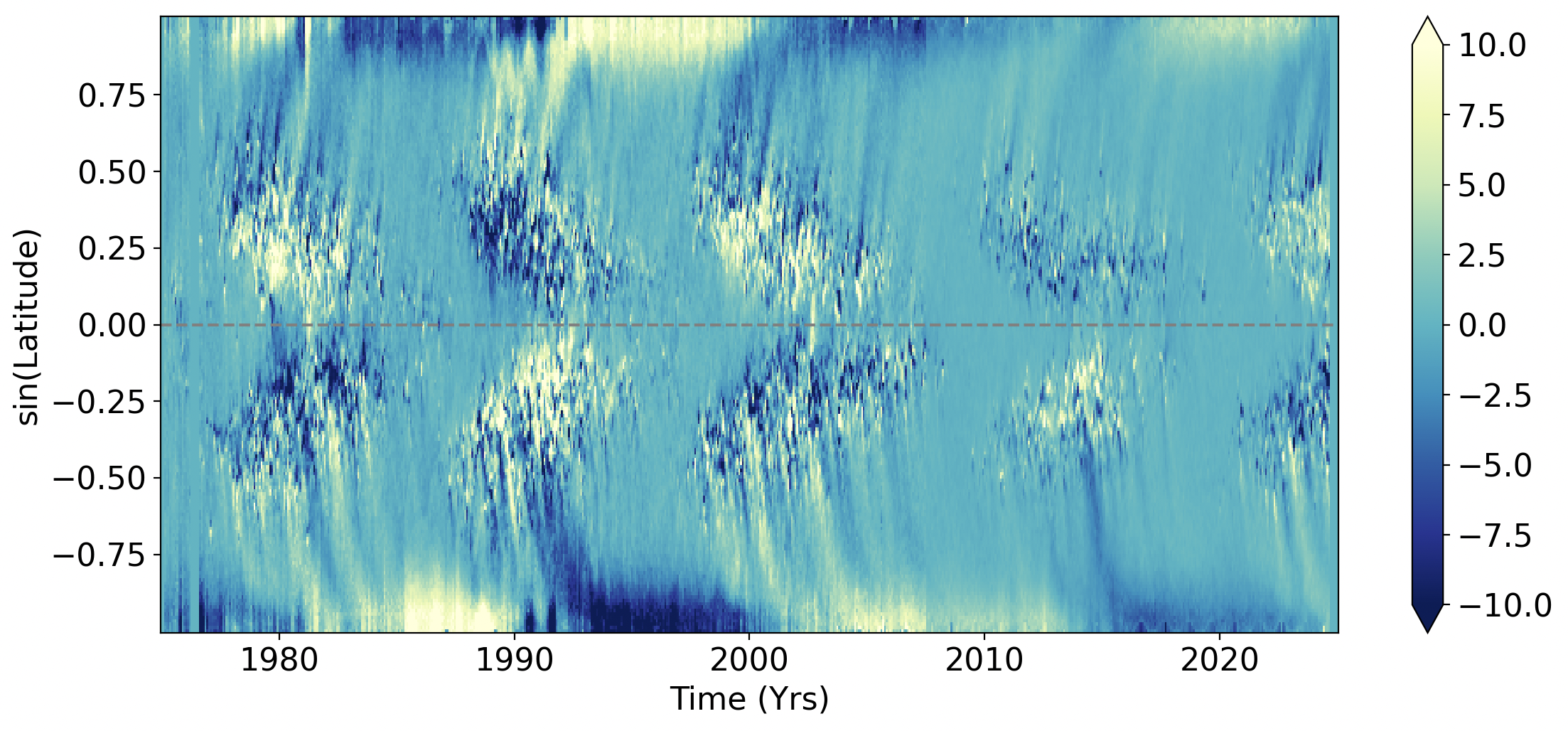}
\includegraphics[width=0.48\textwidth]{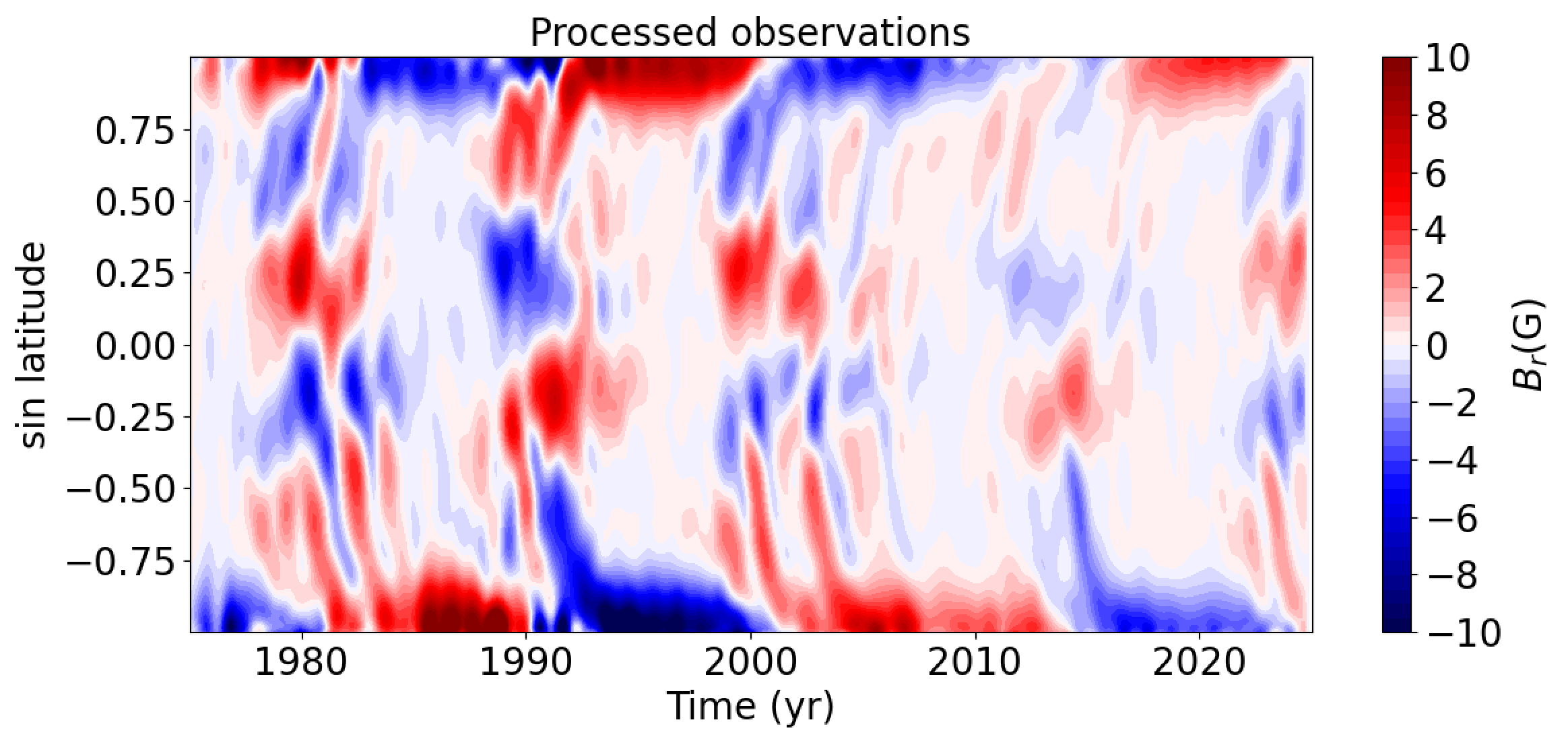}
\includegraphics[width=0.48\textwidth]{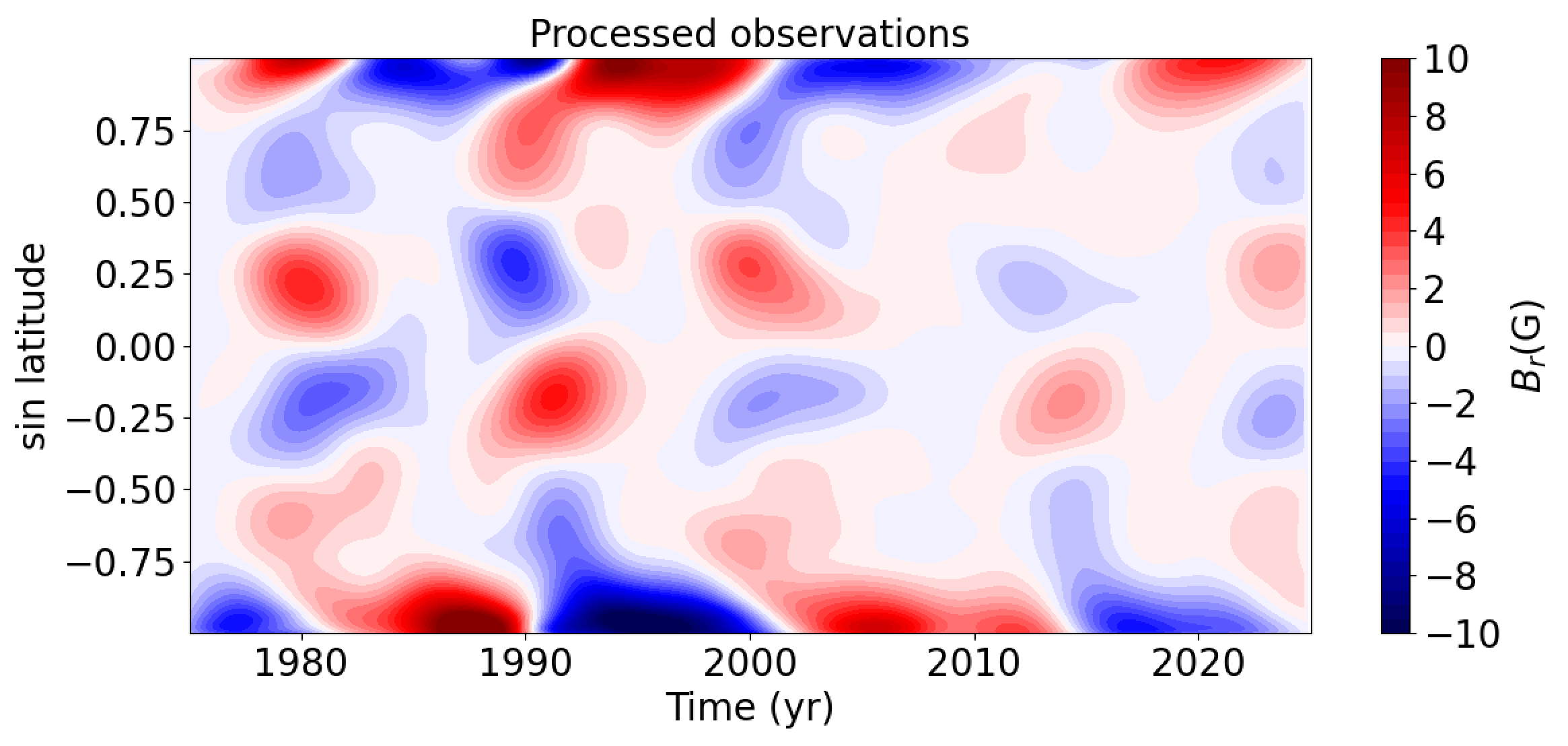}
\end{center}
\caption{Magnetic butterfly diagram: azimuthally-averaged magnetic synoptic maps (coming from various sources) for each Carrington rotation from CR1625 until CR2288 and stacked in time. Top row: unfiltered data and 1-yr (middle row) and 5-yr (bottom row) filtered versions. The later two can be used as line-of-sight magnetic field observations in our assimilation pipeline.}

\label{fig:obsbtfy}
\end{figure}

For the line of sight magnetic field we make use of various sources of synoptic maps generated for each Carrington rotation (CR). In the present paper we use magnetic field data covering the period from early 1975 to April 2024 to be compatible with the 13-month smoothed time series, namely from CR1625 to CR2282 even though we have data series reaching CR2288 (hence as late in 2024 as the 1-month mean SSN). Hence, depending on the time period considered the data for $B_{los}$ comes from different sources (instruments).
When available, we use in priority the GONG synoptic maps from CR2059 to CR2288 \citep{1996Sci...272.1284H}. For the oldest time period we use mostly Wilcox observatory synoptic maps (CR1625-CR2007, CR2015-2017, CR2026, CR2035, CR2040-2042) \citep{1977SoPh...54..353S,1991JGG....43S..59H}. Finally in the case of data gaps we also make use of SOLIS synoptic maps (CR2008-2014, CR2018-2025, CR2027-2034, CR2036-2039, CR2043-CR2058, CR2189) \citep{bertello2015solisvsmpolarmagneticfield}.
From these synoptic maps we then homogenise the data set by projecting high resolution maps on the same latitudinal resolution $N_{\theta}^{o}$ as the Wilcox synoptic maps (e.g. 89 latitudinal mesh points). We then average in longitude and create a butterfly diagram as a function of latitude and time of $B_{los}$ (see Figure \ref{fig:obsbtfy}, top panel). 
In practice, we further use temporal and spatial filters of the butterfly diagram, as the underlying dynamo model cannot capture all the small scale field variations. We have performed several combinations and have settled on two main filtering choices.
We fix the maximum degree $\ell_{max}$ used in the assimilated butterfly diagram to 10. For the temporal filter we choose either 1 or 5-year filtering time windows. Results for the forecasting of the next solar cycle can differ depending on the filtered observation files used during the data assimilation part. However, in most cases considered in this study, the forecast maximum amplitude and timing agree reasonably well, e.g. within the statistical error bars.
In Figure \ref{fig:obsbtfy} we illustrate the original data used and its 1-yr and 5-yr temporal filtered versions assuming $\ell_{max}=10$ as latitudinal spatial filter. We see that the main polarity inversions and equatorward and poleward dynamo branches are retained as are the polar flux accumulation and timing of reversal within small acceptable variations. 

\subsection{Data assimilation strategy}

The assimilation setting adopted in this work is similar to the one used in \cite{Hungetal2017ApJ}, except that real solar observations are used here where only synthetic data were assimilated in our previous studies. We only recap the key features here and we refer to \cite{Hungetal2017ApJ} for the detailed description of the method.

\subsubsection{Data assimilation setting}

The idea of the present work is to assimilate data over a certain period of time, often covering a full cycle period and thus a typical total time of about 11 years, to constrain the control vector which will be used to produce a forecast for the future magnetic field trajectory. Our control vector consists of a part related to the meridional flow and another part related to the initial magnetic field. As we do not want to reconstruct the flow and field pointwise because the dimension of the control vector would then be too large, we expand our meridional flow and initial magnetic field on well-chosen basis functions. The coefficients of those expansions constitute our control vector. The flow will thus be determined by the coefficients $d_{k,l}$ as defined in section \ref{sec_model} and the magnetic field will be represented by its expansion on the eigenvectors of the covariance matrix of the dynamo model, as explained in Appendix B of \cite{Hungetal2017ApJ}.

\begin{figure}[h!]
\begin{center}
\includegraphics[width=0.48\textwidth]{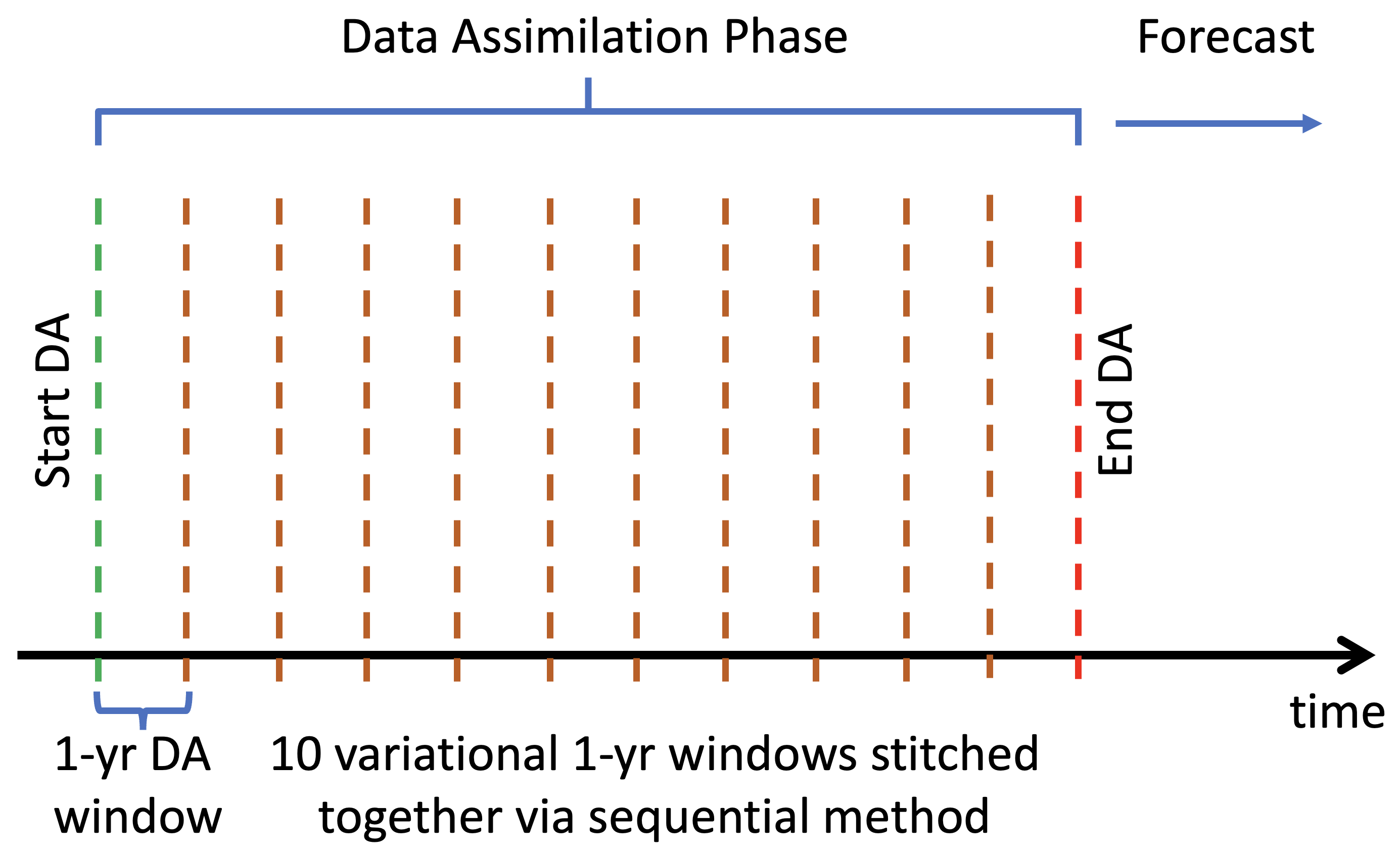}
\end{center}
\caption{Schematic of data assimilation procedure}
\label{fig:da}
\end{figure}

To assimilate data into our dynamo model, we use here a variational assimilation technique, which consists in adjusting a control vector which minimizes a well-defined objective (or cost) function measuring the misfit between outputs of the model and observations. The minimization is performed using a quasi-Newton technique that requires the computation of the function to be minimized and its gradient with respect to each component of the control vector. The choice of the objective function is quite crucial in this procedure and details of each term involved in our function $\mathcal{J}$ are given below, as well as details about the control vector composed of the meridional flow and the initial magnetic field. As we wish that our meridional flow varies in time to produce a best-fitting trajectory for the magnetic field over the whole assimilation window, we decide to use the variational approach on shorter windows (typically of one year) over which our meridional flow is steady. We consequently use a variational technique within each 1-yr window over the 10-yr time span of the data assimilation phase, with each 1-yr window being the initial guess of the next one in a sequential approach. Figure \ref{fig:da} shows a schematic of our assimilation strategy. 

We can summarize our procedure as follows: for the first year of the total assimilation window, the initial guess comes from a dynamo model based on a unicellular flow with a
magnetic cycle of 22 years (i.e. the only non-zero coefficient in the MC expansion is $d_{1,2}=1/3$). Then, for the subsequent assimilation windows, the initial
guess will be the forecasted magnetic field and velocity at the end
of the previous assimilation. Within each one-year window, we estimate
the coefficients of the stream function and initial condition
which minimizes the objective function defined below, and consequently, we obtain a meridional flow described as a piecewise constant function in time.

After this assimilation procedure, our previously described Babcock-Leighton flux transport dynamo model is run to produce the forecasted trajectory for the magnetic field, using the meridional flow and initial magnetic conditions resulting from the assimilation of the last window.

\subsubsection{Objective function and regularization}
\label{sec_obj}

As stated above, the idea of the variational method is, within each window, to minimize a well defined objective function which measures the misfit between the observed quantities and the corresponding outputs from the model.
 The objective function we define in this work is the sum of two terms measuring a misfit to the observations. The first term is the misfit of the line of sight magnetic field:
 
\begin{equation}\label{eq:jlos}
\mathcal{J}_{los}=\sum_{i=1}^{N_t^o} \sum_{j=1}^{N_\theta^o} \frac{\left[B_{los}(\theta_j,t_i)-B_{los}^o(\theta_j,t_i)\right]^2}%
{\sigma^2_{B_{los}}(\theta_j)},
\end{equation}

Here $B_{los}$ is the surface radial field predicted by our mean field dynamo model and $B_{los}^o$ is the observation extracted from the butterfly diagram. The weight $\sigma_{B_{los}(\theta)}$ is simply $1/\sin(\theta)$. This additional factor enables to put more weight on the activity belt at mid-latitudes, as this is where most sunspot appear.

The second term is related to the misfit in the SSN in the Northern and Southern hemispheres, 

\begin{equation}\label{eq:jw}
\mathcal{J}_{W}=\sum_{i=1}^{N_t^o}  \frac{\left[W_{N}(t_i)-W_{N}^o(t_i)\right]^2}{\sigma^2_{W_N}}+ \frac{\left[W_{S}(t_i)-W_{S}^o(t_i)\right]^2}{\sigma^2_{W_S}},%
\end{equation}

where ${\sigma_{W_N}}$ and ${\sigma_{W_S}}$ represent the uncertainties on respectively the Northern and Southern SSN measurements. 
As stated above, the SSN in the Northern and Southern hemispheres have been observed and monitored in various solar stations since the mid 20th century and SILSO provides hemispherical SSN series since 1992 (note that we extended back in time until 1975 using the UCLE station). However, in our dynamo model, we do not produce spots at the surface of our domain. We thus have to use a proxy for the SSN, to be compared to the observables. As in \cite{Hungetal2017ApJ}, this proxy is taken to be the toroidal field at the base of the convection zone, integrated over a small radial extent and averaged in latitude. We thus define $W_N(t)$ and $W_S(t)$ our northern and southern hemisphere sunspot proxies as:
 \begin{eqnarray}
      W_N(t)&=fac \, \times&\int_{\theta=0}^{\theta=\pi/2} \int_{r=0.7}^{r=0.71} B^2_{\phi}(r,\theta,t) r^2 \sin \theta dr d\theta \\
      W_S(t)&=fac\, \times&\int_{\theta=\pi/2}^{\theta=\pi} \int_{r=0.7}^{r=0.71} B^2_{\phi}(r,\theta,t) r^2 \sin \theta dr d\theta
 \end{eqnarray}

where $fac$ is a scaling factor so that our proxies are directly comparable to the observed SSN values. The value of $fac=150$ is used here. 

In this work, we also impose some constraints and regularization on our control vector: the meridional flow and initial magnetic conditions. As described above, we adopt a decomposition of the stream function on Legendre polynomials in latitude and on sine functions in radius. We thus control the meridional flow by imposing some constraints on the coefficients of this expansion. In particular, we choose to:
\begin{itemize}
\item impose that the only non-zero coefficients in the meridional flow expansion are $d_{1,2}$ (representing the 1-cell per hemisphere flow) and $d_{1,3}$ (representing a flow with 3 cells in latitude, to account for some degree of North/South asymmetry in the flow). 
\item impose that the initial magnetic field of one assimilation window stays close to the value of the magnetic field at the end of the previous window. This enables to prevent discontinuities in the magnetic field between successive windows and to propagate information from the previous assimilation to the next one. The first additional term in the objective function then reads:

\begin{equation}\label{eq:j1}
   \mathcal{J}_1=  \sum_{i=0}^{Nmag} \left(c(i)-c_{old}(i)\right)^2
\end{equation}

where $Nmag=10$ is the number of eigenfunctions used to expand our magnetic initial conditions and $c(i)$ is the coefficient on the $ith$ eigenfunction in the current 1-yr window and  $c_{old}(i)$ is the coefficient on the $ith$ eigenfunction in the previous 1-yr window \citep[see][for further details about our expansion procedure]{Hungetal2017ApJ}.

\item impose that $d_{1,3}$ should be less than a certain percentage $p$ of $d_{1,2}$. We adopt the value of $p=10\%$ in this work. This adds the following term:

\begin{equation}\label{eq:j2}
    \mathcal{J}_2=  \sum_{k=1}^{m} \sum_{l=1}^n \left(\max(0, \vert d_{k,l}\vert - p \times \,\vert d_{1,2}\vert )\right)^2 \,\,\, \mbox{for} \,\,\, d_{k,l} \,\,\, \neq d_{1,2}
\end{equation}

\item impose that the deviation from the value of $d_{1,2}^0=1/3$ for $d_{1,2}$ (producing a period of around 22 years) is less than a certain amount $p_2$. We adopt values between $p_2=1\%$ and $10\%$ in this work. This term reads:

\begin{equation}\label{eq:j3}
    \mathcal{J}_3= \left(\max(0, \vert d_{1,2}-d_{1,2}^0 \vert - p_2 \times \, d_{1,2}^0)\right)^2
\end{equation}

\end{itemize}

The assimilation procedure thus consists in the minimization of the full objective function $\mathcal{J}$ composed of the sum of all the terms listed above in equations \ref{eq:jlos}, \ref{eq:jw}, \ref{eq:j1}, \ref{eq:j2} and \ref{eq:j3}, weighted by the different coefficients $A_{los}, A_W, A_1, A_2$ and $A_3$ : 

\begin{equation}
    \mathcal{J}=A_W \mathcal{J}_W+ A_{los}\mathcal{J}_{los} + A_1 \mathcal{J}_1+ A_2 \mathcal{J}_2+ A_3 \mathcal{J}_3
\end{equation}

We note that we choose to fix $A_{los}=1$ so that the relative importance of the SSN observations compared to the line-of-sight radial field is solely determined by the value of $A_W$.
An investigation of the best choices for the regularization coefficients $A_2$ and $A_3$ compared to the value of the misfit coefficients $A_W$ and $A_{los}$ was conducted on the prediction of cycle 24 and the results are presented in section \ref{sec_param}. 
The last coefficient $A_1$ is chosen in order to maintain a satisfying degree of continuity between 2 successive 1-yr windows and to impose some memory between the ten 1-yr windows in the assimilation procedure, ensuring that additional information is gained by assimilating each successive window. It is fixed to a value of $A_1=10^2$ for all cases discussed below. We note here that coefficients $A_1$, $A_2$ and $A_3$ are for the moment not part of our control vector as we wanted to focus on the physical quantities (i.e. the meridional flow and magnetic field initial conditions). Assimilating those extra three parameters in our pipeline is postponed to later studies.

Our assimilation procedure finally consists in adjusting a control vector constructed by the 2 meridional flow coefficients $d_{1,2}$ and $d_{1,3}$ and 10 coefficients representing the projection of the magnetic field on the eigenvectors of the covariance matrix. Our assimilated data consist in 89 latitudinal points times 12 months for $B_{los}$ and 1 value for each hemisphere times 12 months for the monthly-smoothed SSN. We thus have a largely overdetermined problem, in which removing part of the data would thus have a limited impact on our assimilation results.

\subsubsection{Ensemble members}

We here describe how we produce an ensemble of predictions for the next solar cycle when the data assimilation procedure described above is over.

Creating an ensemble of forecast trajectories is a common procedure as it allows to test the sensitivity to initial conditions of the prediction. It is well known since \cite{Lorenz63} that small perturbations in a dynamical system can lead to large variations on a long time scale.
We have shown in \cite{Hungetal2017ApJ} that our data assimilation pipeline, when let free to evolve beyond its final assimilated observations data points, takes between 10 to 15 years to reach the same level of errors/mismatch in a controlled experiment than a free dynamo run.
Hence the temporal horizon limit for our forecast is of the order of one solar cycle. Beyond that 10 or 15-yr horizon limit, the uncertainty has grown too large to be able to consider the prediction as meaningful. 

\begin{figure}[h!]
\begin{center}
\includegraphics[width=0.48\textwidth]{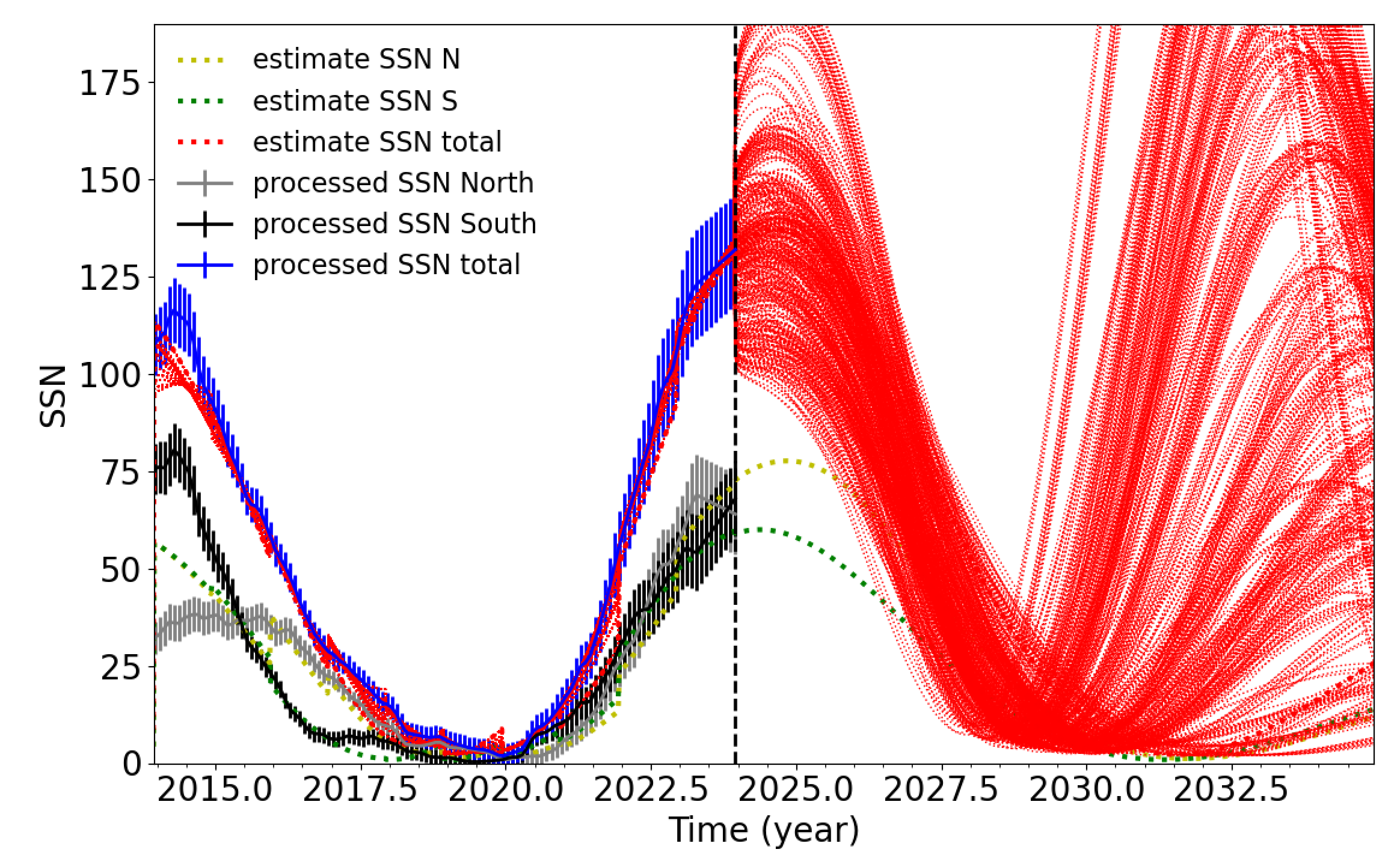}
\end{center}
\caption{Trajectory of the 480 ensemble members. Example of hemispherical data assimilation starting in mid 2013 and ending in $\tau_{exyr} =$ mid 2023. The $12 \times 40 = 480$ extrapolations (forecast ensemble trajectories) computed by the solar dynamo model are starting beyond that time. We clearly see that further we look in the future larger the spread is both in amplitude and timing of the next maximum of cycle 26. Forecast of the next minimum around 2030 is within our temporal horizon limit as the time of the writing of the paper. In order not to overcomplicate the figure, we voluntarily omit to plot the 2 $\times$ 480 trajectories of each hemisphere.}
\label{fig_members}
\end{figure}

In order to probe this uncertainty and its growth, it is convenient to generate an ensemble of trajectories by perturbing the initial magnetic field conditions and the meridional circulation used to generate the forecast trajectory. To do so, we introduce a random perturbation to the central values obtained at the end of the assimilation procedure. The time at which we introduce this random perturbation is called the extrapolation time $\tau_{exyr}$. Before that point, the dynamo model is being controlled via our data assimilation procedure to follow the solar activity data and beyond that point, it is let free to go with the initial conditions and meridional flow obtained during the assimilation and a fixed set of parameters.
We can choose the number of ensemble members $N_{ens}$. We usually use values between 30 and 60 ensemble members for each value of $A_{W}$, which allows us to sample reasonably well the most probable forecast trajectories for a given set of parameters and initial magnetic conditions. We further use a set of 12 $A_W$ values $A_W=[0.2,0.5,0.8,0.9,1,2,3,6,11,21,51,101]$, from loosely tracking the SSN vs tightly following it during the assimilation phase. Overall, if we choose 40 ensemble members per $A_W$ values, we get a final set of 480 distinct trajectories (members) that constitutes our global ensemble.
In Figure \ref{fig_members}, we show each of the 480 trajectories in the most general hemispherical version of the pipeline for illustration purposes. Please note that these trajectories are obtained by perturbing both the meridional flow and the magnetic field, explaining the sharp but continuous change of the 480 trajectories with respect to the mean seen in the figure. The level of perturbations have been chosen such as to provide a representative sampling of the observed historical variations of the 11-yr solar cycle.

\subsubsection{Total vs hemispherical}
As stated before, our data assimilation pipeline can assimilate either the total (global) SSN time series or its hemispherical counterpart. The only difference is the objective function where we either consider the two hemispheres separately or together. The objective function shown in Eq. \ref{eq:jw} distinguishes the two hemispheres. When they are assimilated together, function $\mathcal{J}_W$ becomes:
\begin{equation}\label{eq:jwt}
\mathcal{J}_{Wtot}=\sum_{i=1}^{N_t^o} \frac{\left[W_{T}(t_i)-W_{T}^o(t_i)\right]^2}{\sigma^2_{W_T}},%
\end{equation}
with ${\sigma_{W_T}}$ the uncertainty on each SSN measurement, $W_{T}^o(t_i)=W_{N}^o(t_i)+W_{S}^o(t_i)$ the total observed SSN and $W_{T}(t_i)=W_{N}(t_i)+W_{S}(t_i)$ the proxy for the total SSN defined as
\begin{equation}
      W_T(t)=\int_{\theta=0}^{\theta=\pi} \int_{r=0.7}^{r=0.71} B_{\phi}^2(r,\theta,t) r^2 \sin \theta dr d\theta
 \end{equation}

\section{Validation of the forecasting strategy on cycles 22, 23 and 24}
\label{sec_validation}

This first result section is dedicated to addressing the quality of our assimilation pipeline on the 3 previous cycles namely 22, 23 and 24 for which the true SSN evolution is known. In the next section we will focus on the ongoing cycle 25 for which an estimate of the maximum as well as the next minimum is given.

\subsection{Forecasting of the past 3 cycles}

\subsubsection{Overall prediction as a function of the cycle phase}
In this section, we wish to assess the quality of our data assimilation and forecasting procedures on past cycles for which the evolution of the sunspot numbers in both hemispheres are known. Our main goal here was to test under which conditions our technique would produce accurate and precise predictions for the timing and maximum of cycle $n+1$, when data was assimilated on cycle $n$ at different phases of the cycle. We noticed that an important parameter was the weight $A_W$ put on the observations of the SSN and thus decided to vary this parameter to produce an ensemble from which predictions are estimated.

For the assimilation step, we use 6 different values for $A_W$, namely $1, 2, 6, 11, 21$ and $51$, except for cycle 22 where we added smaller $A_W$ values of $0.2, 0.5, 0.8, 0.9$. The total assimilation window is chosen to be 10 years in length, divided into 10 one-year windows. As stated before, we test the effect of assimilating different phases of the cycle. We thus have 5 different cases, with the total assimilation window ending before the solar minimum, then around the solar minimum, and then at 1 year, 2 years and 3 years after the minimum. For example, in order to forecast cycle 24, we use windows covering cycle 23 in the intervals $[1997,2007]$, $[1998,2008]$, $[1999,2009]$, $[2000,2010]$  and $[2001,2011]$. 

After the assimilation procedure, we get a best-fit model from which we calculate the predicted trajectories. To do so, the meridional flow coefficients obtained after the data assimilation for each model are perturbed using a Gaussian noise with a typical amplitude of $6\times 10^{-5}$ to produce an ensemble of 40 members per value of $A_W$ (cf. Figure \ref{fig_members}). A beam of predictions is then produced for each cycle phase, either for each $A_W$ or for all the values of $A_W$ together.

\begin{figure}[h!]
\begin{center}
\includegraphics[width=0.48\textwidth]{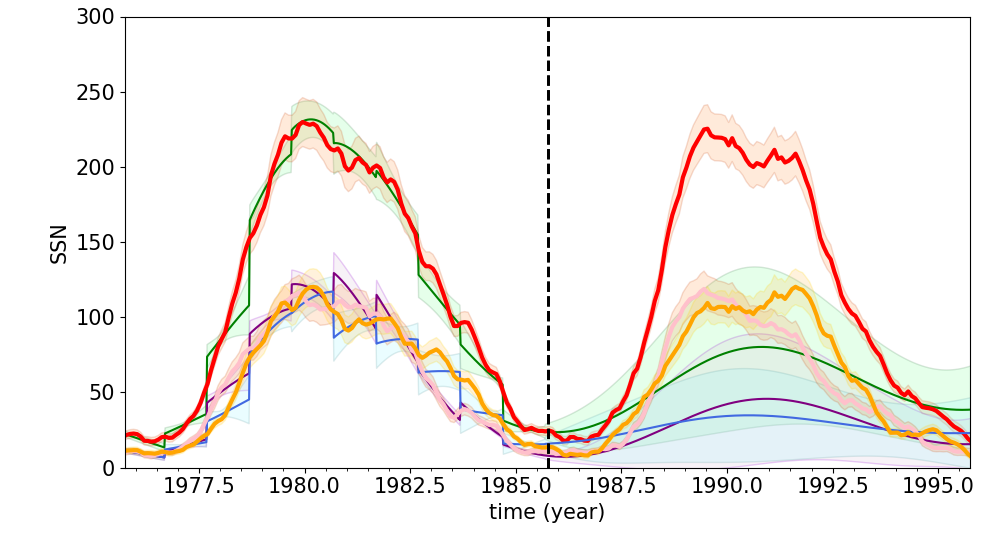}
\includegraphics[width=0.48\textwidth]{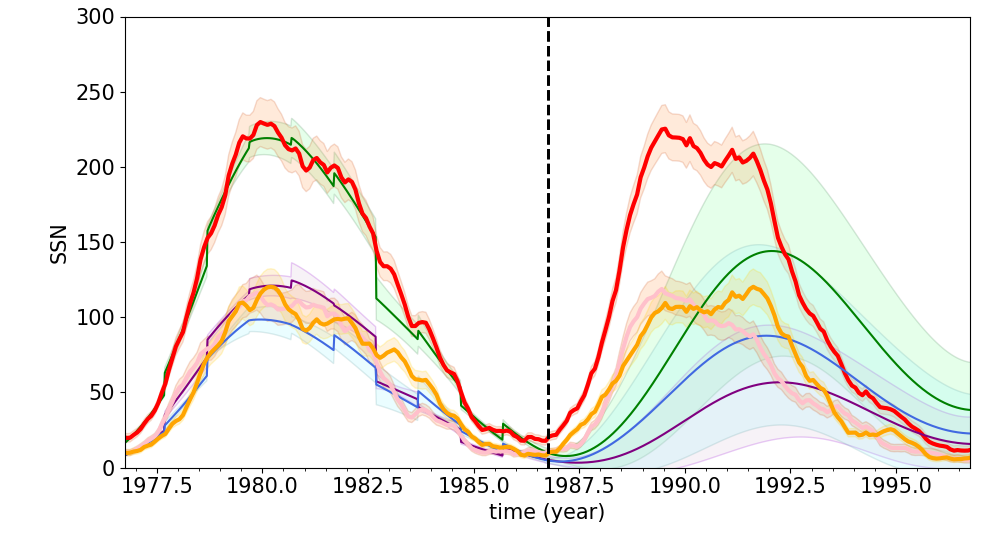}
\includegraphics[width=0.48\textwidth]{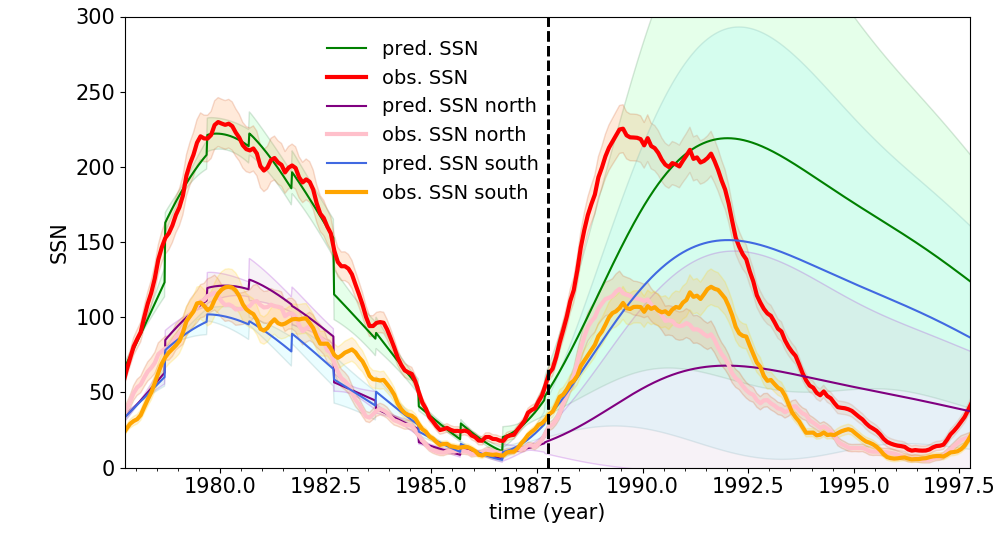}
\includegraphics[width=0.48\textwidth]{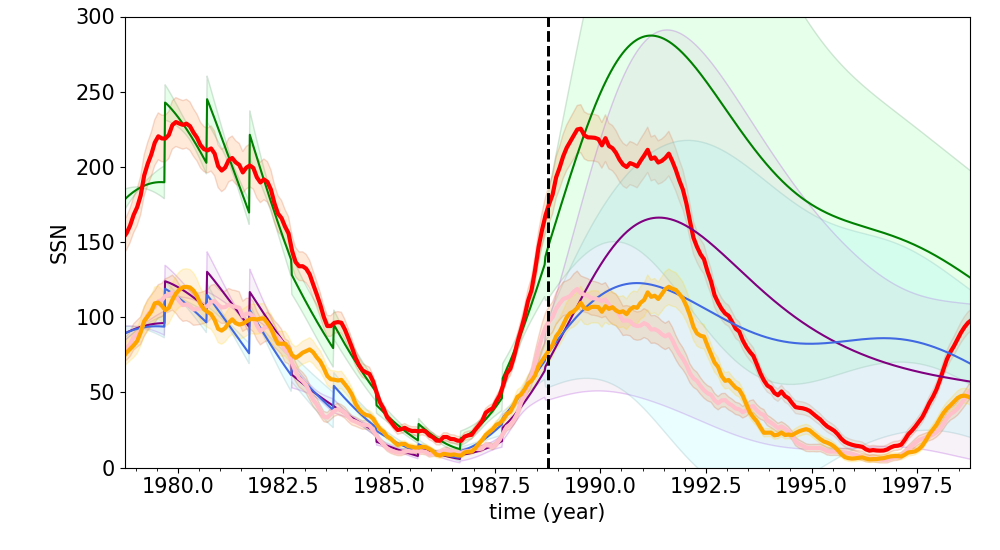}
\includegraphics[width=0.48\textwidth]{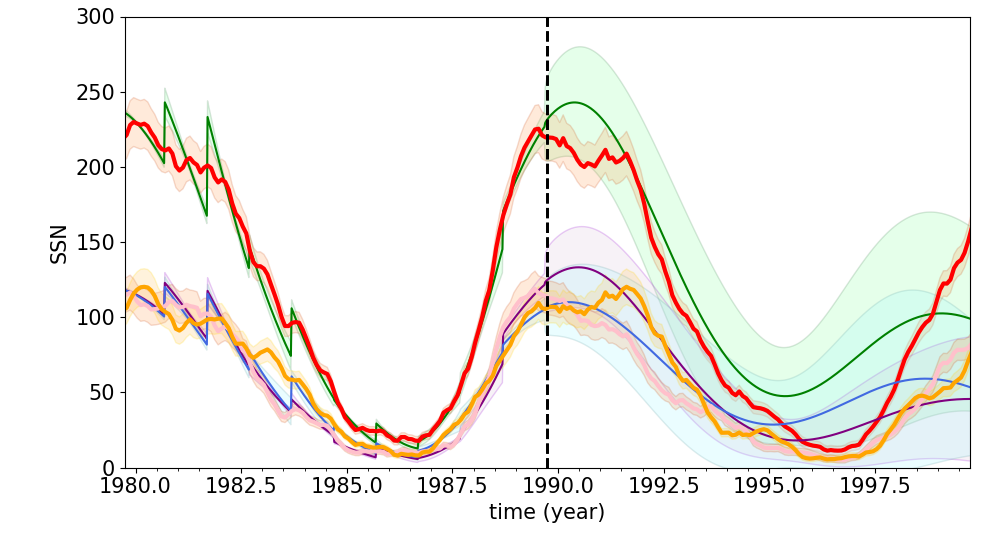}
\end{center}
\caption{Prediction for the sunspot number of cycle 22 as a function of time, at different assimilated cycle phases. For the sake of clarity, the legend (same for all plots) is indicated only on the 3rd panel.}
\label{fig_cycle22}
\end{figure}

\begin{figure}[h!]
\begin{center}
\includegraphics[width=0.48\textwidth]{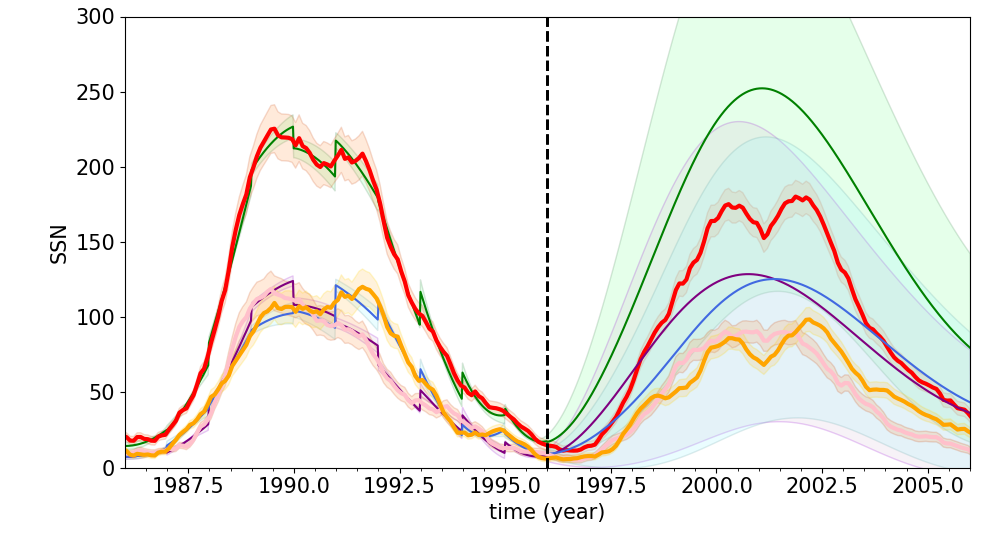}
\includegraphics[width=0.48\textwidth]{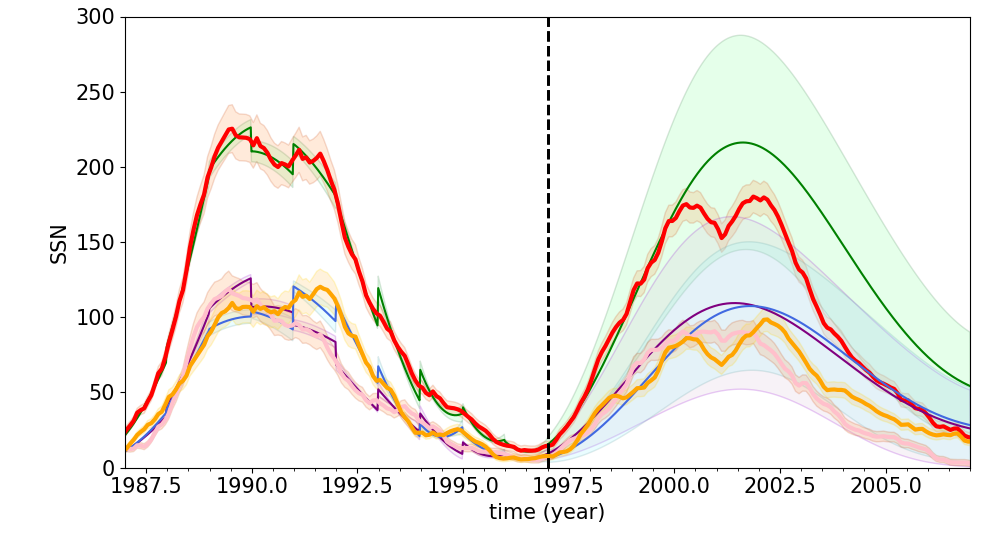}
\includegraphics[width=0.48\textwidth]{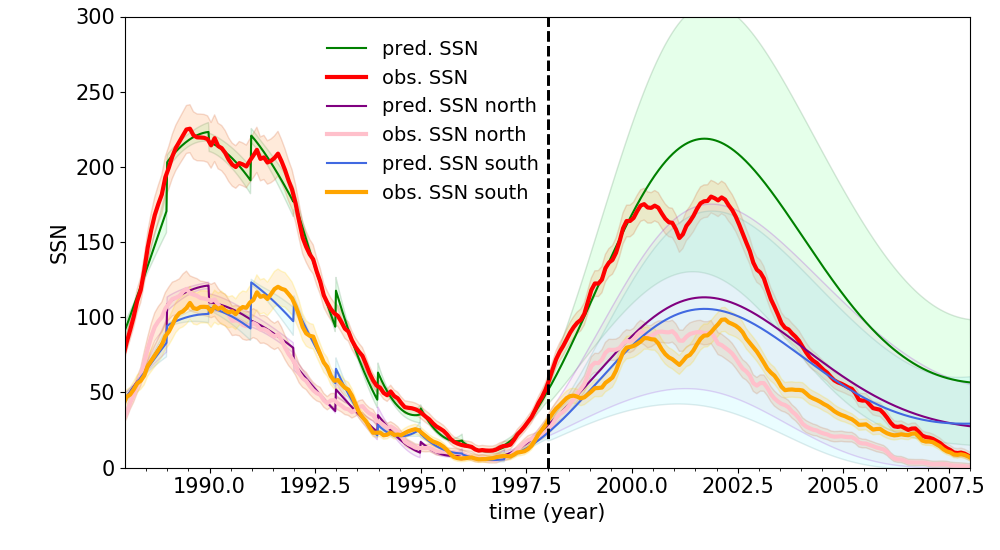}
\includegraphics[width=0.48\textwidth]{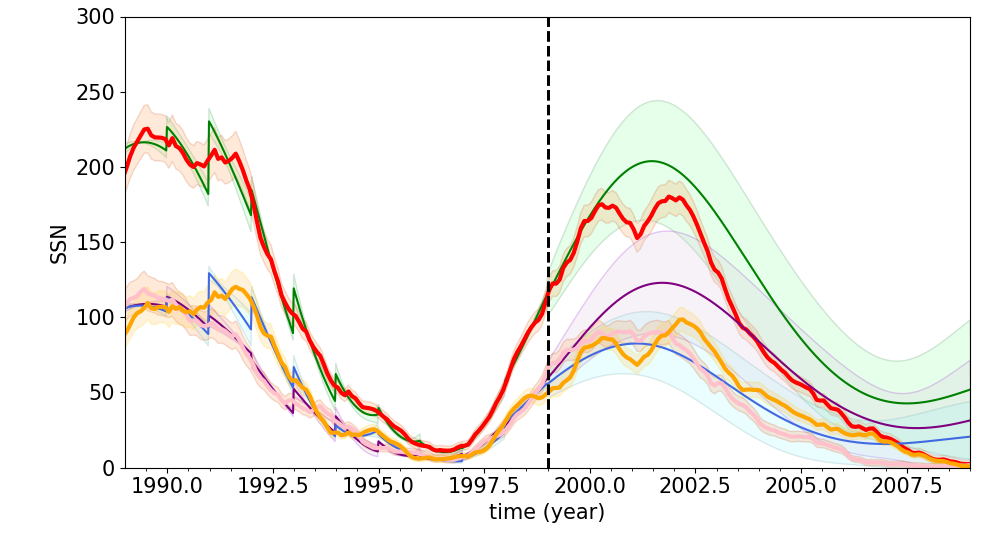}
\includegraphics[width=0.48\textwidth]{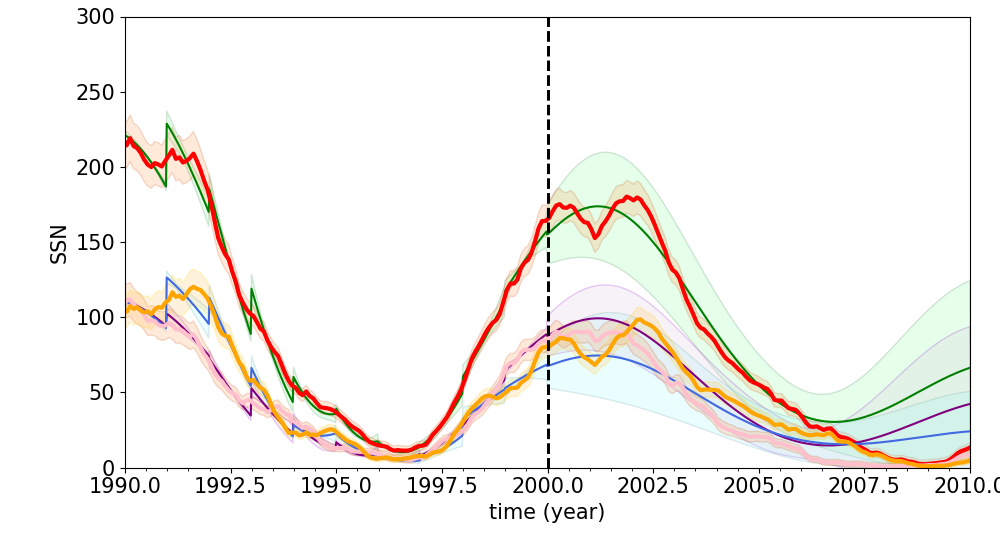}
\end{center}
\caption{Same as Figure \ref{fig_cycle22} but for cycle 23.}
\label{fig_cycle23}
\end{figure}

\begin{figure}[h!]
\begin{center}
\includegraphics[width=0.48\textwidth]{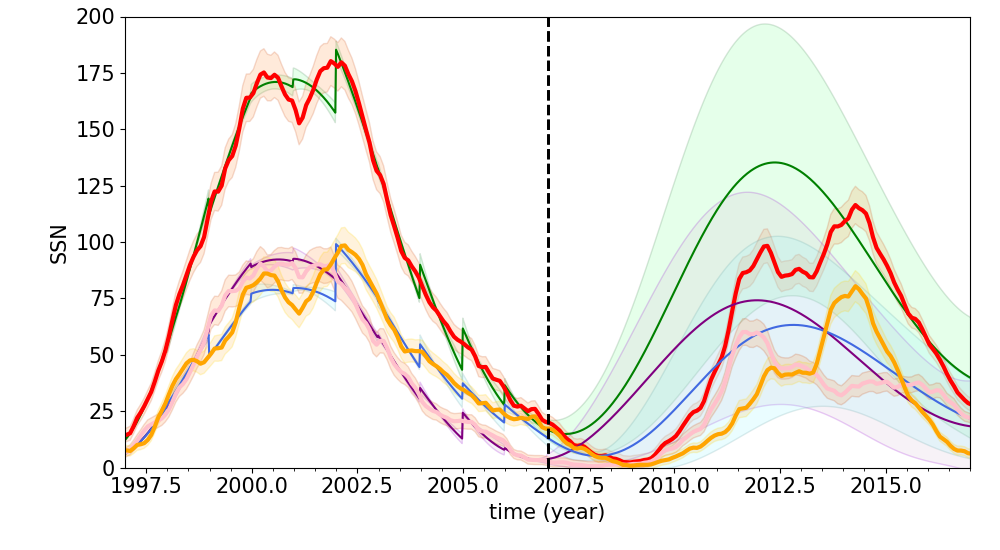}
\includegraphics[width=0.48\textwidth]{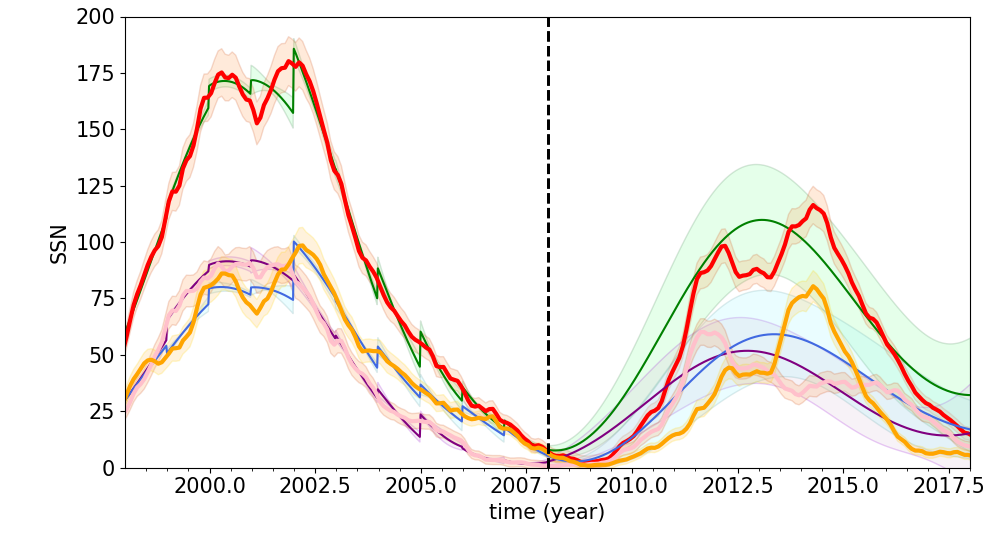}
\includegraphics[width=0.48\textwidth]{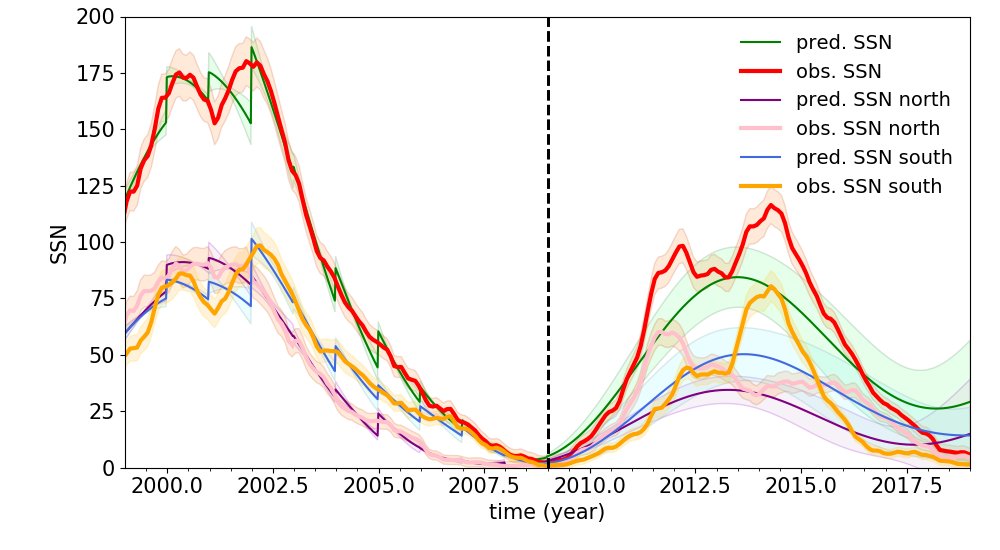}
\includegraphics[width=0.48\textwidth]{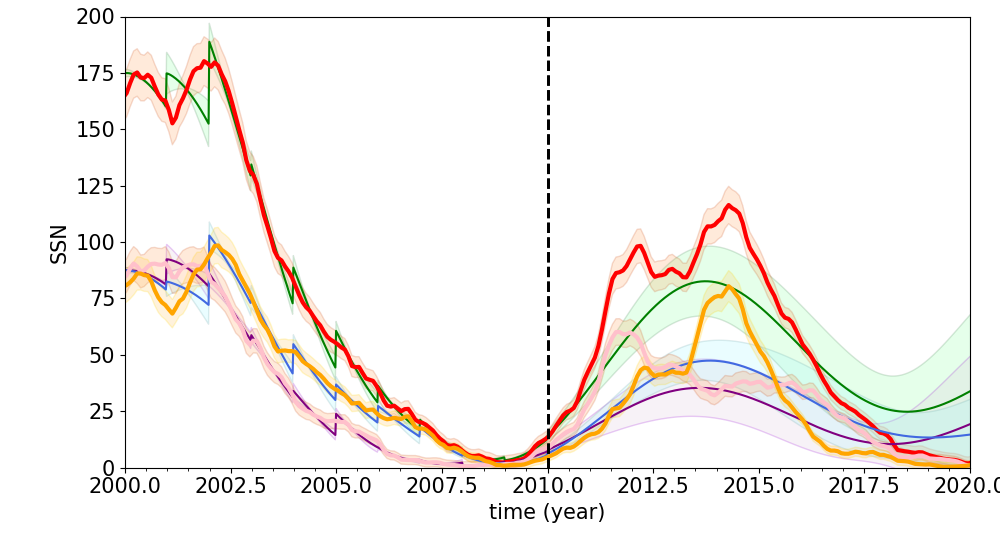}
\includegraphics[width=0.48\textwidth]{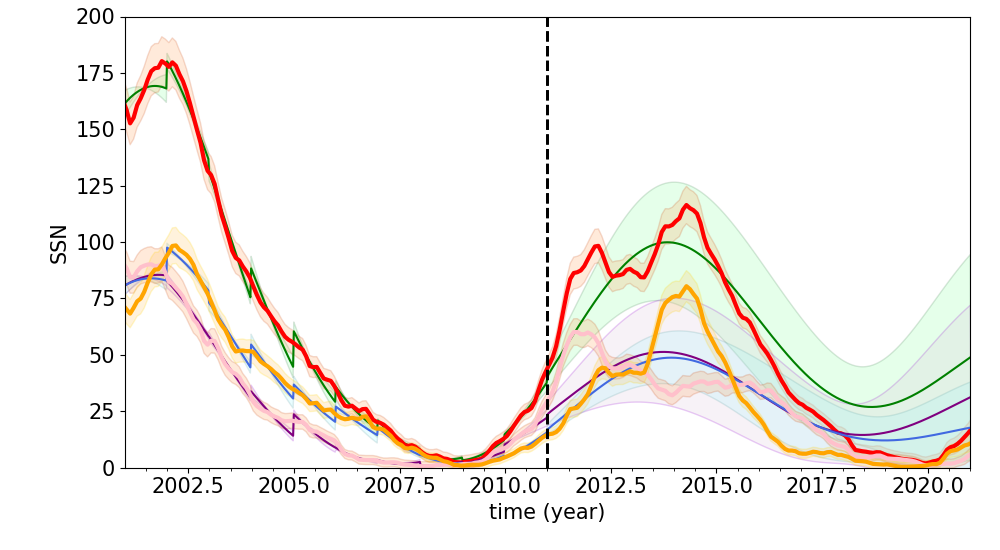}
\end{center}
\caption{Same as figs \ref{fig_cycle22} and \ref{fig_cycle23} but for cycle 24.}
\label{fig_cycle24}
\end{figure}

Results for cycles 22, 23 and 24 are illustrated in Figures \ref{fig_cycle22}, \ref{fig_cycle23} and \ref{fig_cycle24}. In addition to these figures, more quantitative estimates of the statistics of the distributions of the time at maximum $t_{max}$ and of the sunspot number at this maximum $\rm SSN_{max}$ are given in the appendix in table \ref{tab}. Those more quantitative results are further illustrated in Figure \ref{fig_boxes}.

Let us first focus on Figures \ref{fig_cycle22},\ref{fig_cycle23} and \ref{fig_cycle24}. Those figures show that for all cycles, the assimilation phase (until the vertical dashed line) performs well at following the hemispherical and total SSN data. Since we apply a variational assimilation on separate 1-yr windows, we see some discontinuities between consecutive windows due to the fact that continuity between windows is constrained on the initial magnetic conditions (by the term weighted by $A_1$ in the objective function, see Equation \ref{eq:j1}) but not on the meridional flow coefficients. The forecasting part is of course more interesting to focus on. The green curve indicates the total predicted SSN while the purple (resp. blue) indicates the predicted SSN in the Northern (resp. Southern) hemisphere. The true values are shown in red, pink and yellow respectively. In all the predicted curves, the 1-$\sigma$ dispersion is shown in shaded areas around the mean curves. The date at which the assimilation phase ends is here clearly an important parameter in the success of the prediction. As expected, forecasts initiated in the rising phase usually perform better at following the evolution (even after the maximum) than forecasts starting slightly before the previous minimum. This is particularly clear when we compare the last panels of each figure, where the correct evolution of the total and hemispherical data is usually captured up to 4 to 5 years after the maximum (i.e. close to the next minimum) and the first panels where the predictions most of the time fail right away at following the true curves. In between, the forecasted curves get closer and closer to the real values, with smaller and smaller error bars. This is true mostly for cycles 23 and 24. Cycle 22 seems to present more difficulties, in particular the error bars on the second and third panels are large, owing to the fact that 2 different populations of predictions (weak and strong cycles) appear for different values of the weighting parameter of the SSN values $A_W$. We come back to cycle 22 in section \ref{sec_cycle22}.

 We note that the plots show the results of the predictions using all the values of $A_W$ together. However, we noticed in our procedure that small values of $A_W$ perform better for a prediction made from the previous cycle minimum, while larger values of $A_W$ perform better when the predicted cycle has already started its rising phase. This can be understood by the fact that $A_W$ represents the weight given to the SSN (represented by the integral of $B_\phi$ at the tachocline in our model) compared to the weight given to the surface radial field. At solar minimum, there will of course be only a few sunspots, while the strength of the radial field will be the main progenitor for the sunspot-producing toroidal field of the next cycle \citep{2023SSRv..219...40B}.

We see that our procedure enables to get asymmetries between the Northern and Southern hemispheres, the blue and purple curves of Figures \ref{fig_cycle22}, \ref{fig_cycle23} and \ref{fig_cycle24} peaking at different values and at different times. 
If we focus on the maximum SSN values, we see that the level of asymmetry produced in our model is compatible with what is seen in the observations. Indeed, the relative difference between the Northern and Southern SSN typically reaches $10$ to $30\%$ in the observations as well as in the predicted curves. However, what our pipeline has more difficulty to reproduce is the surge-like behaviour of the yellow and pink curves of Figures \ref{fig_cycle23} and \ref{fig_cycle24}, which produce very distinct peaks in the Northern and Southern hemispheres. This surge-like behaviour is indeed not designed to be reproduced in our current pipeline where the forecast trajectories coming from our dynamo model are close to sinusoidal and thus much smoother. If we now focus on the timing of the cycle, we see that the model does not produce asymmetries as pronounced as in the real Sun. This is especially true for cycle 24 where the 13-month smoothed data show a 2.6 years delay between both maxima. In comparison, our model predicts only a maximum delay of less than 1 year. This could be due to the fact that we constrained in our procedure the only coefficient producing an asymmetry with respect to the equator, namely $d_{1,3}$, to be smaller than $10\%$ of $d_{1,2}$. Relaxing this constraint and adding more coefficients to the meridional flow would probably help to produce less similar behaviours for the North and South hemispheres and possibly higher delays between the maxima.

\subsubsection{Convergence of predictions}
\begin{figure}[h!]
\begin{center}
\includegraphics[width=0.48\textwidth]{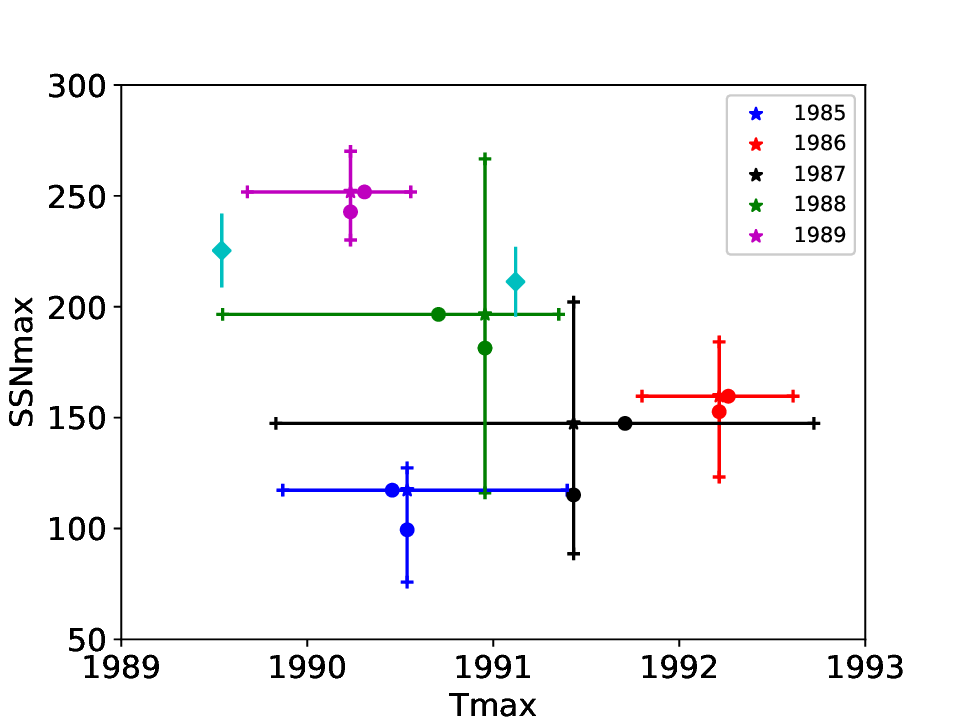}
\includegraphics[width=0.48\textwidth]{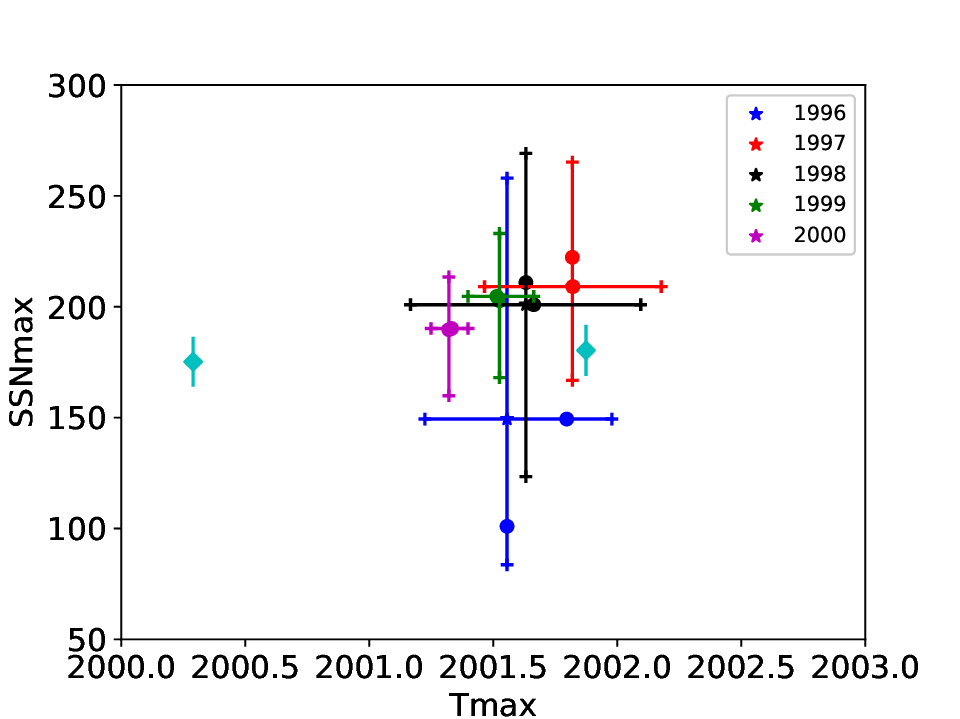}
\includegraphics[width=0.48\textwidth]{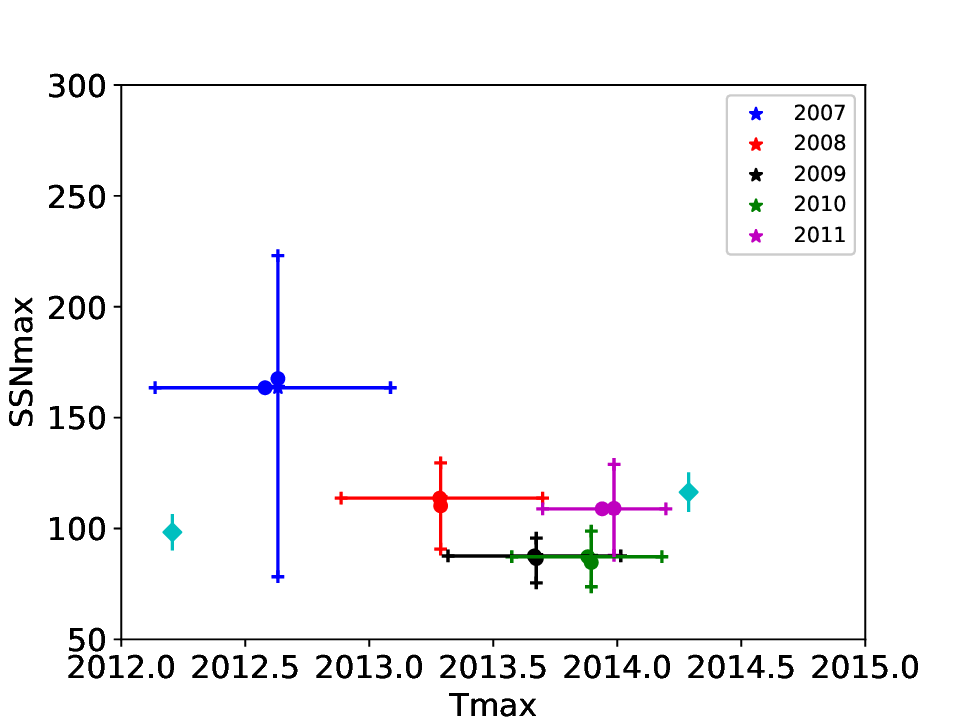}
\end{center}
\caption{Summary of the predicted time and sunspot number at maximum for cycles 22 (upper panel), 23 (mid panel) and 24 (lower panel) predicted from different phases of the previous cycle. The error bars are taken between the 1st and 3rd quartiles of the distributions of $t_{max}$ and $\rm SSN_{max}$ given in table \ref{tab}. The 2nd quartiles (i.e. median values) are indicated as dots and mean values as stars located at the intersection of the error bars. The cyan dots indicate the true values of the maximum with error bars on the sunspot number. For all cycles, 2 cyan dots are present, representing the 2 peaks of similar amplitude in the different cycles, visible on the 13-month smoothed sunspot number (red curves of Figures \ref{fig_cycle22}, \ref{fig_cycle23} and \ref{fig_cycle24}). }
\label{fig_boxes}
\end{figure}

Let us now turn to Figure \ref{fig_boxes}, where we show the prediction only of the maximum total SSN and the time of this maximum for the different cycle phases shown in the previous figures. To produce these plots, we calculated the statistics (mean, median, 1st and 3rd quartiles) of the distribution of the points of coordinates $(t_{max}, \rm SSN_{max})$ for all members of the ensemble. We note that the values of the means in this figure will be different from the maximum values of the mean curves shown in Figures \ref{fig_cycle22}
, \ref{fig_cycle23} and \ref{fig_cycle24} and thus should not be directly compared.
Moreover, for these more quantitative estimates, some selection of cases has been applied. To be considered in our statistics, a predicted maximum had to happen before 7 years after the end of the assimilation phase and be contained in the range $[60,350]$, i.e. consistent with minimum and maximum values ever observed in the past solar cycles. The extent of the horizontal and vertical lines are determined by the 1st and 3rd quartiles of the distributions. The mean values are found at the intersection between these two lines. The cyan diamonds show the true values of the maximum SSN. We usually plot two values since the cycles usually present a double peak due to the delay between the hemispheres. This is quite obvious when looking at the red curves of Figures \ref{fig_cycle22}, \ref{fig_cycle23} and \ref{fig_cycle24}.

Figure \ref{fig_boxes} clearly shows the convergence of our predictions while the assimilated cycle phase gets closer to the maximum. In all cases, the error bars indeed decrease and the predicted values get closer to the true values. We note that our predictions do not show a double-peak shape since the hemispheres remain quite in phase with each other. We thus estimate that the prediction could be considered as successful when a value between the two true peaks is reached.  We also note from Figure \ref{fig_boxes} that the errors made on the timing of the predicted cycle is in general much less (in percentage) than the error made on the sunspot maximum. For cycle 23, the maximum error is around 1 year and reduced to about 6 months for cycle 24. This could be due to the fact that our background dynamo model is calibrated to produce an 11-yr cycle while no a priori is introduced for the sunspot number. 

\subsubsection{Difficulties with Cycle 22?}
\label{sec_cycle22}

The previous section showed that the predictions of Cycle 22 gave significantly less good results than for the 2 other cycles. Indeed, as seen in Figure \ref{fig_cycle22}, the predictions starting from the previous minimum (in 1985 and 1986) systematically give a very weak amplitude of the next maximum compared to the true value. When the assimilated period extends past the minimum phase (i.e. ends in 1987 and 1988), the error bars on the predicted future (both the maximum SSN and timing) are quite large. For example, for the prediction starting in 1987, the difference between the 3rd and 1st quartiles is about $100\%$ of the median while the same value reduces to $70\%$ for the same cycle phase for cycle 23 and to $30\%$ for cycle 24. The error bars are relatively small as can be seen in the bottom panel of Figure \ref{fig_boxes}. The reason for the larger dispersion in the predictions for cycle 22 is unclear but possible explanations may be given. First, it is quite obvious that this cycle was stronger than the 2 other ones and the prediction from the previous minimum may be more difficult if too much weight is put on the high $A_W$ values. This is why we included for this cycle additional values of $A_W<1$, such that more weight is put on the radial line-of-sight magnetic field compared to the SSN. We indeed find that low values of $A_W$ perform in general better at capturing the high maximum of cycle 22 compared to higher values. 

\begin{figure}[h!]
\begin{center}
\includegraphics[width=0.48\textwidth]{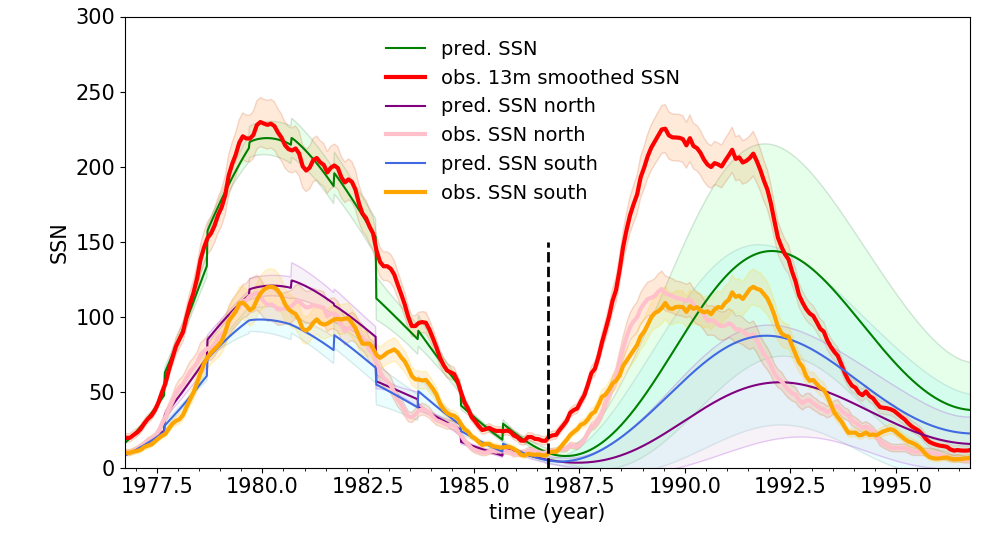}
\includegraphics[width=0.48\textwidth]{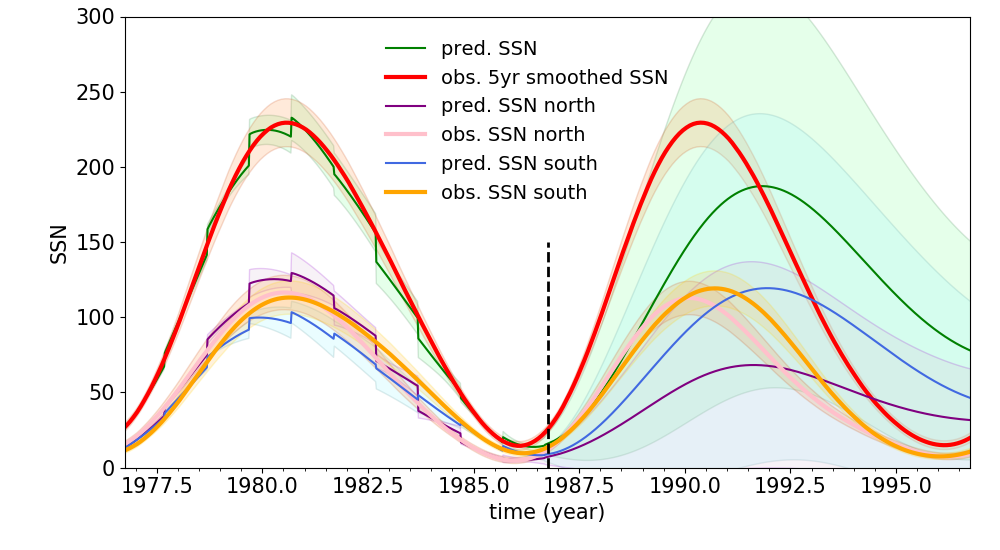}
\end{center}
\caption{Predictions for cycle 22 starting from year 1986, using the 13-month smoothed SSN data (top) and the 5-yr smoothed time series (bottom).}
\label{fig_smooth}
\end{figure}

A second possibility for the peculiarity of this cycle is the long-lasting minimum which can be seen in Figure \ref{fig_cycle22}, especially for the hemispherical data. We indeed observe that the curves remain very flat around the same value of the SSN of around 10 to 15 in each hemisphere for almost 2 years between 1985 and 1987. Especially for an assimilation phase finishing within this period (first and second panel of Figure \ref{fig_cycle22}), the model seems to have difficulties catching the particularly fast rise of the next cycle. This lingering minimum is quite obvious on the 13-month smoothed SSN data but less clear if the 5-yr smoothed curves are used. Incorporating the 5-yr smoothed data in our assimilation and forecast procedures, we indeed find quite different results. The results are illustrated in Figure \ref{fig_smooth} where the predictions are shown starting from year 1986 using the 13-month smoothed (top panel) and 5-yr smoothed (bottom panel) SSN time series. All other parameters are kept identical between the 2 cases. It is clear from this figure that the sharper minimum of the 5-yr smoothed curve allows the model to better catch the rising phase of the cycle. The rising phase being better reproduced, the prediction for the next maximum is thus closer to the true value. 

This test shows that it may be worth considering various filtering windows for the SSN as well as for the butterfly diagram in an ensemble prediction to minimize the adverse effect of some peculiar features, like the long-lasting minimum appearing mostly in the 13-month smoothed data here, to impact and bias the forecast.

\subsection{Influence of parameters}
\label{sec_param}

In this subsection, we justify our choice of the weighting parameters $A_i$'s of the background terms involved in the objective function $\mathcal{J}$. We particularly focus on the values of $A_2$ and $A_3$ which respectively appear in front of the term imposing that the coefficients of the meridional circulation other than $d_{1,2}$ are not higher than $10\%$ of $d_{1,2}$ and in front of the term ensuring that $d_{1,2}$ stays close to its nominal value of $1/3$. 

In order to test the effect of these two coefficients, we fix the other coefficients to the following values: $A_w=11$, $A_{los}=1$ and $A_1=10^2$ while varying $A_2$ and $A_3$ from $10^3$ to $10^6$. We assess the quality of the prediction by measuring the distance between the predicted maximum (both its amplitude $\rm SSN_{max}$ and its timing $t_{max}$) and the true maximum available for cycles 23 and 24. 

\begin{figure}[h!]
\begin{center}
\includegraphics[width=0.48\textwidth]{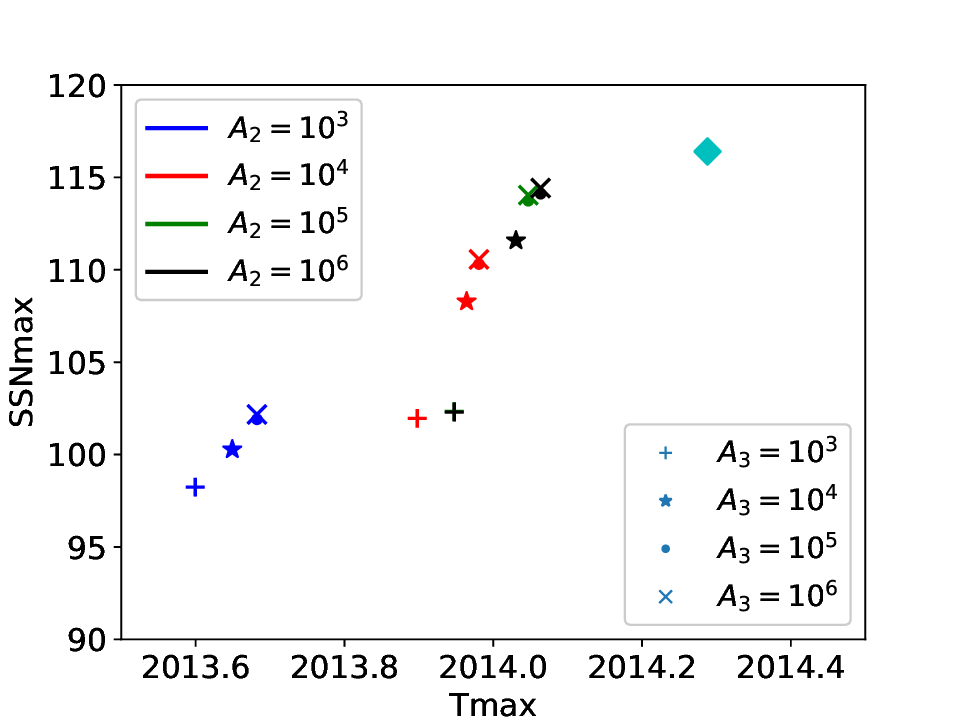}
\end{center}
\caption{Predictions of the value and timing of the maximum sunspot number for cycle 24 for different values of $A_2$ and $A_3$. The included cases correspond to a 10-yr assimilation window extending until 2011 for the prediction of cycle 24. The various symbols correspond to different values of $A_2$ (colors) and $A_3$ (markers). The cyan symbol is the true maximum.}
\label{fig_param}
\end{figure}

The results are presented in Figure \ref{fig_param} for the prediction of cycle 24. It is rather clear from this figure that some cases perform poorly for the prediction of the maximum. For example, the blue and red colors corresponding to values of $A_2=10^3$ and $10^4$ usually give large misfits, independently of the value of $A_3$. Similarly, plus and stars symbols, corresponding to values of $A_3=10^3$ and $10^4$ usually fail at forecasting the max. We conclude from these parametric surveys that a value of $A_2$ more than $10^4$ and $A_3$ more than $10^5$ are necessary to ensure convergence towards a better forecast. We indeed observe in the figure that beyond those values, the predicted maximum does not change very much and the quality of the convergence thus stalls. In the results presented in this work, we fixed $A_2=10^5$ and $A_3=10^6$, corresponding to the green crosses in the plots, which produce satisfying results for cycle 24. Equivalent tests were conducted on cycle 23 (not shown here) with similar results.

\section{Cycle 25 forecasting}
\label{sec_cycle25}

We now present how our forecasting tool performs for the current cycle 25.
At the time of writing of the paper we are close to the peak of cycle 25. NOAA/NASA consensus in 2019, which the current pipeline was one of the forecasting tools taken into account in drawing the final consensus, was announcing a mean maximum SSN value of 113$\pm 18$ that would be reached between 2023 and 2026. A recent update in October 2023 only based on the Space Weather Prediction Center (\href{https://www.swpc.noaa.gov/}{SWPC}) tools gives a higher SSN value 153$\pm 8$ and a more precise maximum occurring on September 2024 $\pm$ 5 months.

A beta version of our pipeline was giving a low value of 92 $\pm 10$ for the cycle maximum and a timing in mid-2024 \citep{Brun20}. In the next subsection, we present the released version of our 11-yr solar cycle forecasting tool with the objective function described in section \ref{sec_obj} that provides a better control of systematics and forecasted value of the previous cycles 22 to 24 (see section \ref{sec_validation}). We did a rerun of the December 2018 conditions and compared it to various forecasting done since, namely October 2023 as in the latest update cited above and April 2024 for the latest available data. 

\subsection{No cast prediction}
Before discussing what our tool predicts for solar cycle 25, we would like to discuss the {\it no cast} solution of cycle 25. The no cast solution consists of using historical data as predictors. The simplest one is to take the previous cycle 24 or 23 (to respect an odd/even alternation) and to use it to forecast cycle 25. A slightly more elaborate one is to use the mean of either the first 24 cycles or say the last 5. 

\begin{figure}[h!]
\begin{center}
\includegraphics[width=0.48\textwidth]{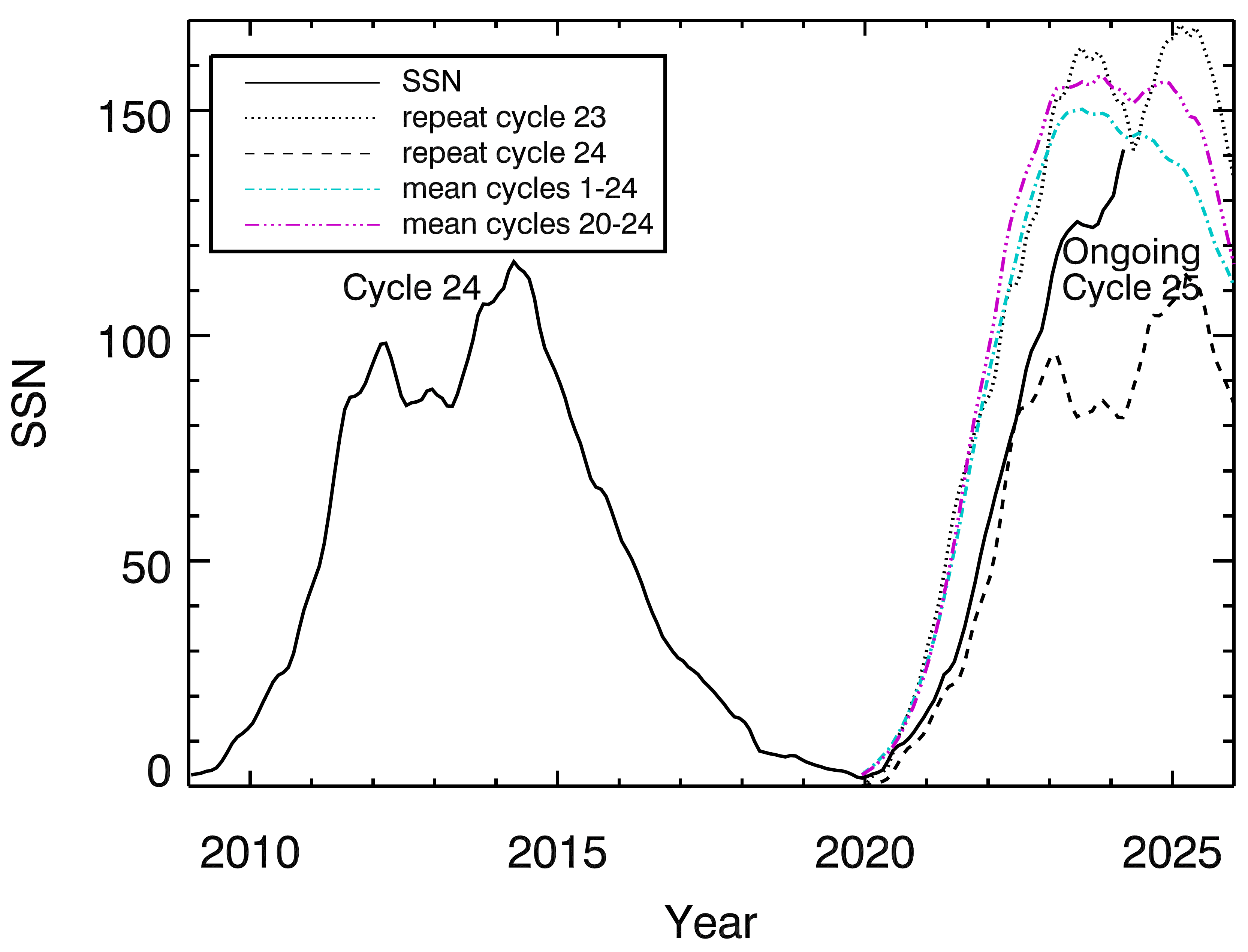}
\end{center}
\caption{No cast of cycle 25: We compare cycle 25 to various historical data: the previous cycle 24 (dashed line) , cycle 23 (the odd numbered previous cycle in dotted line) and two averages of the past record, a mean of the previous 24 cycles (cyan dash-dotted) and of the last 5 (pink dash-dotted). As we can see none of these no cast/historical data are providing an accurate prediction of the ongoing cycle 25.}
\label{fig:nocast}
\end{figure}

As can be seen in Figure \ref{fig:nocast}, this procedure of using previous cycles to predict the next does not provide a very accurate estimate of cycle 25. The two averages are much higher than cycle 25, meaning that this cycle is moderate. Likewise, cycle 24 is not providing a good estimate as it is clearly too weak and by opposition cycle 23 is too strong. This clearly demonstrates how difficult it is to forecast precisely the next solar cycle \citep{Bushby2007ApJ,Petrovay2020}. Historical data are useful as they provide a global probable envelope of what the next solar cycle could be, but it is not accurate enough. Hence, more sophisticated tools such as \emph{solar predict} are needed if we wish to go beyond historical data, see for instance the reviews of  \cite{Hathaway2015LRSP, Petrovay2020} for a list of the various methodologies developed in the community.

Finally one could also consider the timing of the current cycle with respect to the Gleissberg cycle. We recall that the Gleissberg cycle is a possible 90 to 100-yr modulation of the 11-yr cycle envelope. For cycle 25 this would mean considering the amplitude of cycles 14 to 16 (i.e. about 100 years ago) for instance. However, that does not work so straightforwardly as cycles 14 to 16 do not represent a good match to cycle 25. 

\subsection{Prediction of the total SSN}
\label{sec:cyc25tot}

\begin{figure}
\begin{center}
\includegraphics[width=0.48\textwidth]{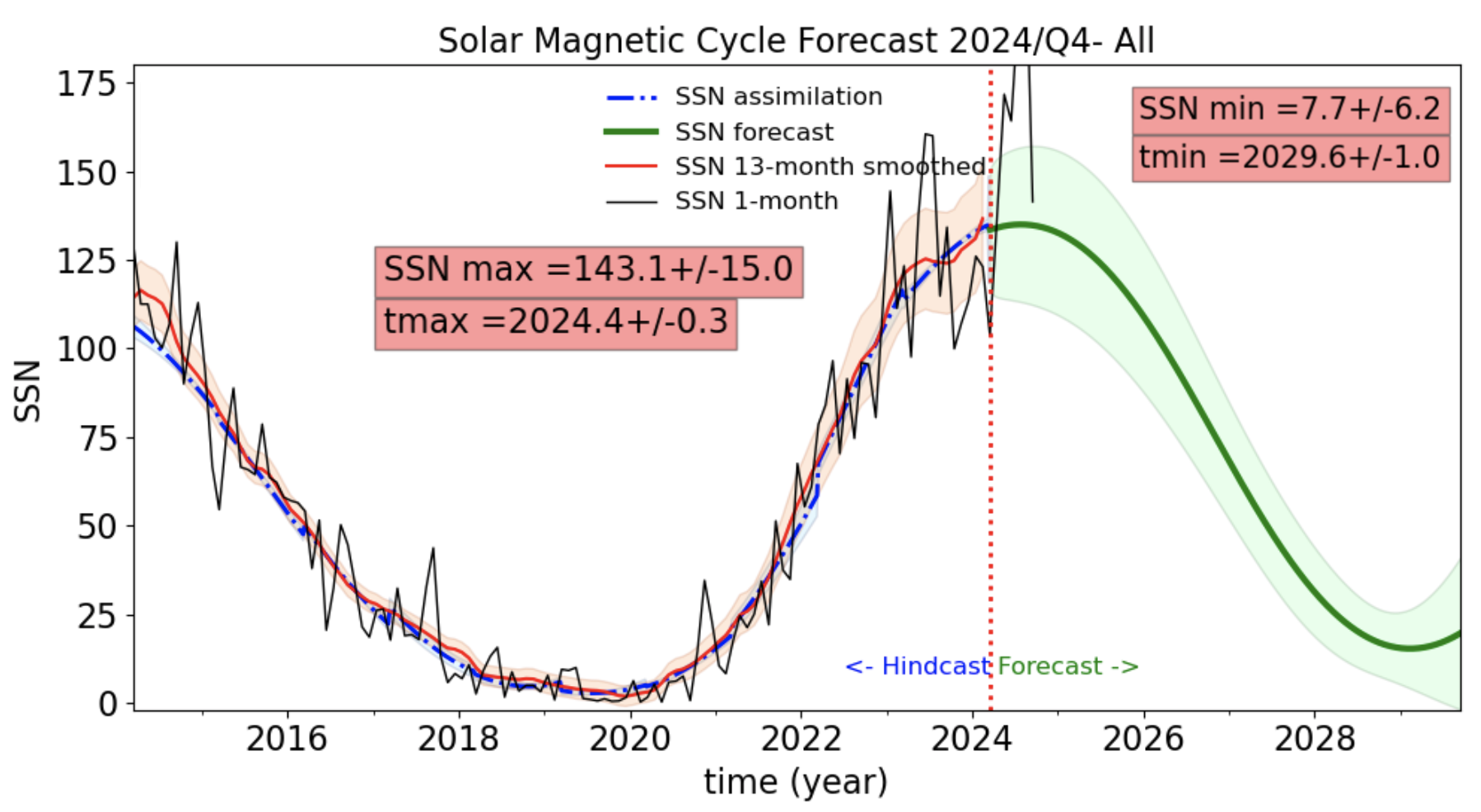}
\end{center}
\caption{Solar cycle 25 forecast as of April 2024, for the sunspot number evolution. The next maximum as well as next minimum values are shown for the total SSN.}
\label{Fig:cycle25tot}
\end{figure}

\begin{figure}
\begin{center}
\includegraphics[width=0.48\textwidth]{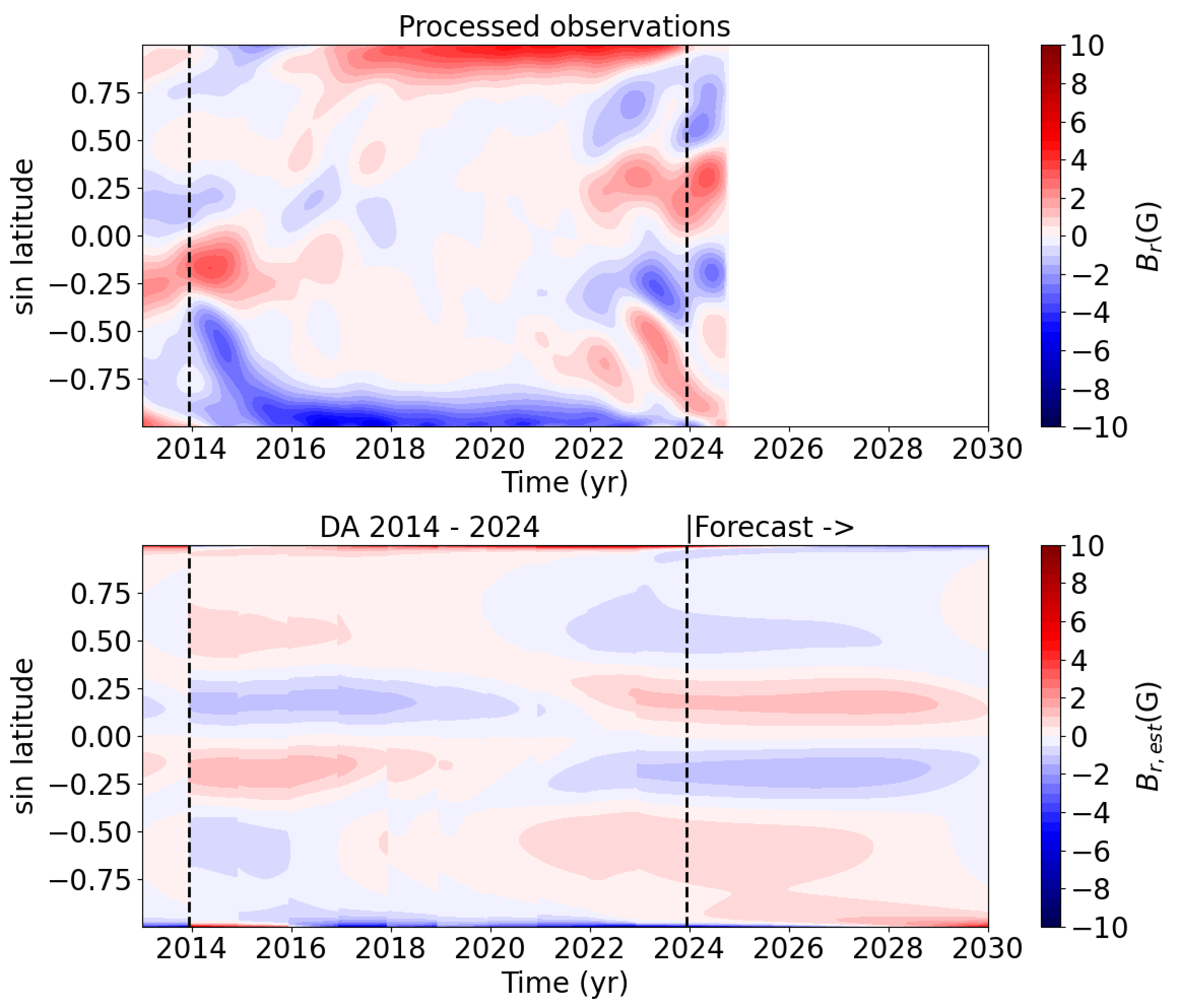}
\end{center}
\caption{Solar cycle 25 forecasts as of December 2023, for the butterfly diagram. The observed radial field is shown in the top panel and the results of the assimilation phase (from 2014 to 2024) and forecast (from 2024) are illustrated on the bottom panel.}
\label{Fig:cycle25btfy}
\end{figure}

We will now turn to our prediction of the current cycle 25. The cycle is already well advanced and such a forecasting must be considered as a validation step as we are likely experiencing in 2024 the peak of solar activity.
In Figures \ref{Fig:cycle25tot} and \ref{Fig:cycle25btfy} we show the typical output of our \emph{solar predict forecasting tool}. As explained before, we are assimilating the SSN and the butterfly diagram data into our solar dynamo model. By the means of a minimisation procedure we adapt the meridional circulation coefficients to match the assimilated data over a 10-yr long period. We clearly see that the DA algorithm is doing very well for the SSN evolution (Figure \ref{Fig:cycle25tot}), as the model trajectory (in blue dash triple dot line style) closely follows the 13-month smoothed SSN time series (in red). Note that the small disagreement at the very beginning is solely due to the time it takes in the first variational window to adjust the guess solar dynamo model to the solar data.

The agreement is less clear during the assimilated phase of the butterfly diagram (Figure \ref{Fig:cycle25btfy}). In particular, the assimilation captures only the largest-scale features of the observed pattern and the polar field gets much more confined close to the poles in our procedure than in the observations. This can be explained by the weight we impose in our objective function on some key observables. Indeed, we put more weight in the assimilation on the SSN than on the $B_{los}$ by imposing a large number of $A_W$ in our ensemble to be larger than $1$. Secondly, by using a $\sigma_{B_{los}}(\theta)$ to be equal to $1/\sin(\theta)$, we also put more weight on the low latitudes and thus do not expect as good a match at high latitudes. We see however here that the behaviour at low latitudes is correctly reproduced, with the positive and negative polarities appearing in the correct latitudinal regions around 2022 after the long period of solar minimum. We note that various functions have been implemented for $\sigma_{B_{los}}(\theta)$ to better take into account the polar field but since our main goal was the SSN evolution, we decided to keep $1/\sin(\theta)$ as this case gives the best fit to the SSN during the assimilation phase.

We now turn to the forecast itself, after the DA procedure has converged over the 10 years before time $\tau_{exyr}$ (indicated by the vertical dotted lines in both figures). 
On Figure \ref{Fig:cycle25btfy}, if we focus on the forecast, we see a smooth evolution in time of both polarities and quite symmetric with respect to the equator. As will be discussed in the next section, we indeed do not produce large asymmetries between the two hemispheres in our current forecast of cycle 25. The smooth evolution is due to our dynamo model which produces, within the range of parameters considered here, a nearly sinusoidal shape for the magnetic fields as a function of time and large-scale features in space. It is of course more interesting here to focus on the forecast in the SSN shown on Figure \ref{Fig:cycle25tot}. The 3-yr SSN forecast trajectory (shown in  green) is here extrapolated from the latest adjusted control vector at $\tau_{exyr}$. We indicate in the figure the forecasted maximum and its timing. Error bars represent the 1-$\sigma$ (standard deviation) statistical uncertainty coming from the 480 ensemble members. Because we are near maximum, the error are here relatively small, representing about 2.4 months in timing and about $\pm 15$ in sunspot number.
We will see in section \ref{sec_cycle25phase} that these errors can be larger if the forecast is made starting from an earlier date in the cycle phase. Our algorithm is currently forecasting on average a maximum this Fall 2024 and a moderate solar cycle amplitude of 143.1 $\pm 15.0$.

\subsection{Prediction of the hemispherical SSN}
Because we are assimilating the SSN time series into a dynamo model that covers the full latitudinal range from a co-latitude of 0 up to $\pi$, we do not have to assume any symmetry with respect to the equator. We can hence assimilate the hemispherical SSN time series rather than the total, hence taking into account potential phase delay between the two solar hemispheres. Indeed it is well known that the Sun's quadrupolar mode leads to asynchronous hemispheres by perturbing the dipolar mode that would otherwise impose the two hemispheres to be synchronous \citep{Derosa2012ApJ, Finley2023}. For instance, cycles 23 and 24 had very distinct reversals of the northern and southern poles, with a time lag for the southern hemisphere of about 17 and 27 months respectively, quite significant when compared to the solar cycle length. Hence it is quite important to be able to capture this underlying dynamics between the solar dipole and quadrupole to improve the forecast. 

Furthermore, since we let the algorithm pick in each hemisphere what meridional circulation matches the best the hemispherical SSN time series, we can produce asynchronous northern and southern forecasts. From historical records we know that the poles reversals cannot be delayed from one another by more than a couple of years. This is due to the fact that the solar quadrupolar magnetic field amplitude is on average of the order of $20$ to $25\%$ of the solar dipole, except during the solar maximum, where the quadrupolar mode $\ell=2 \mbox{ , } m=0$ dominates \citep{Derosa2012ApJ, Finley2023}. Note that usually it is better to refer to equatorially symmetric modes by opposition to anti-symmetric ones as during solar global reversals the equatorial dipole $\ell=1 \mbox{ , } m=1$ is also involved. Indeed, in full 3D spherical geometry it is the sum $\ell \mbox{ + } m$ that determines the parity of the magnetic mode, symmetric modes corresponding to an even sum \citep{McFadden91, Derosa2012ApJ}. Since our dynamo model is 2.5D, we only consider axisymmetric modes such that $m=0$ and the parity of the degree $\ell$ directly provides the equatorial symmetry.

\begin{figure}[h!]
\begin{center}
\includegraphics[width=0.48\textwidth]{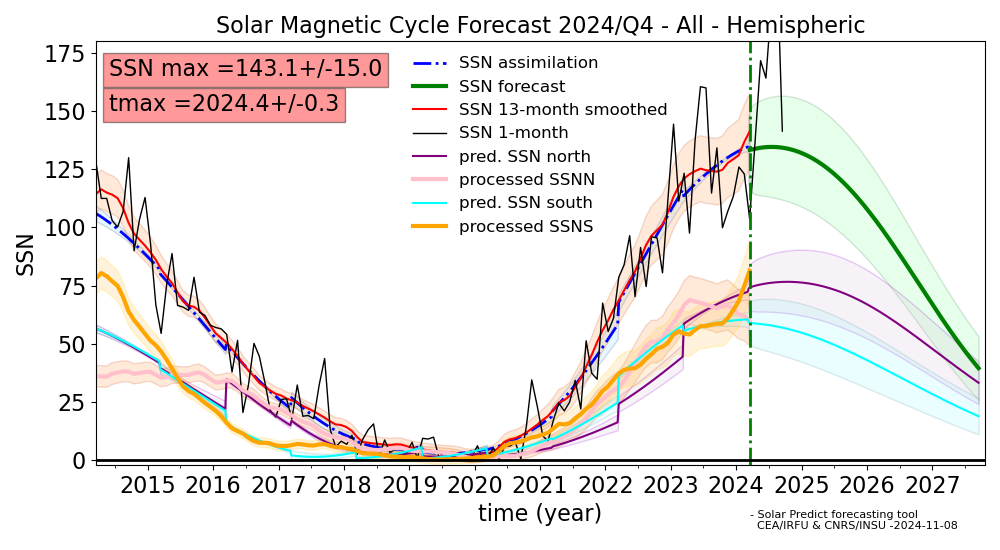}
\end{center}
\caption{Same as Figure \ref{Fig:cycle25tot} but for the hemispherical version of the solar 25 activity forecast.}
\label{Fig:cycle25hemi}
\end{figure}

In Figure \ref{Fig:cycle25hemi} we display the hemispherical forecast of cycle 25. First we note that the DA algorithm is still doing very well at assimilating data in both hemispheres and adapting the control vector in each of them. This is possible thanks to the inclusion of the coefficient $d_{1,3}$ that breaks the equatorial symmetry. 

\begin{figure}[h!]
\hspace{-0.5cm}
\includegraphics[width=0.55\textwidth]{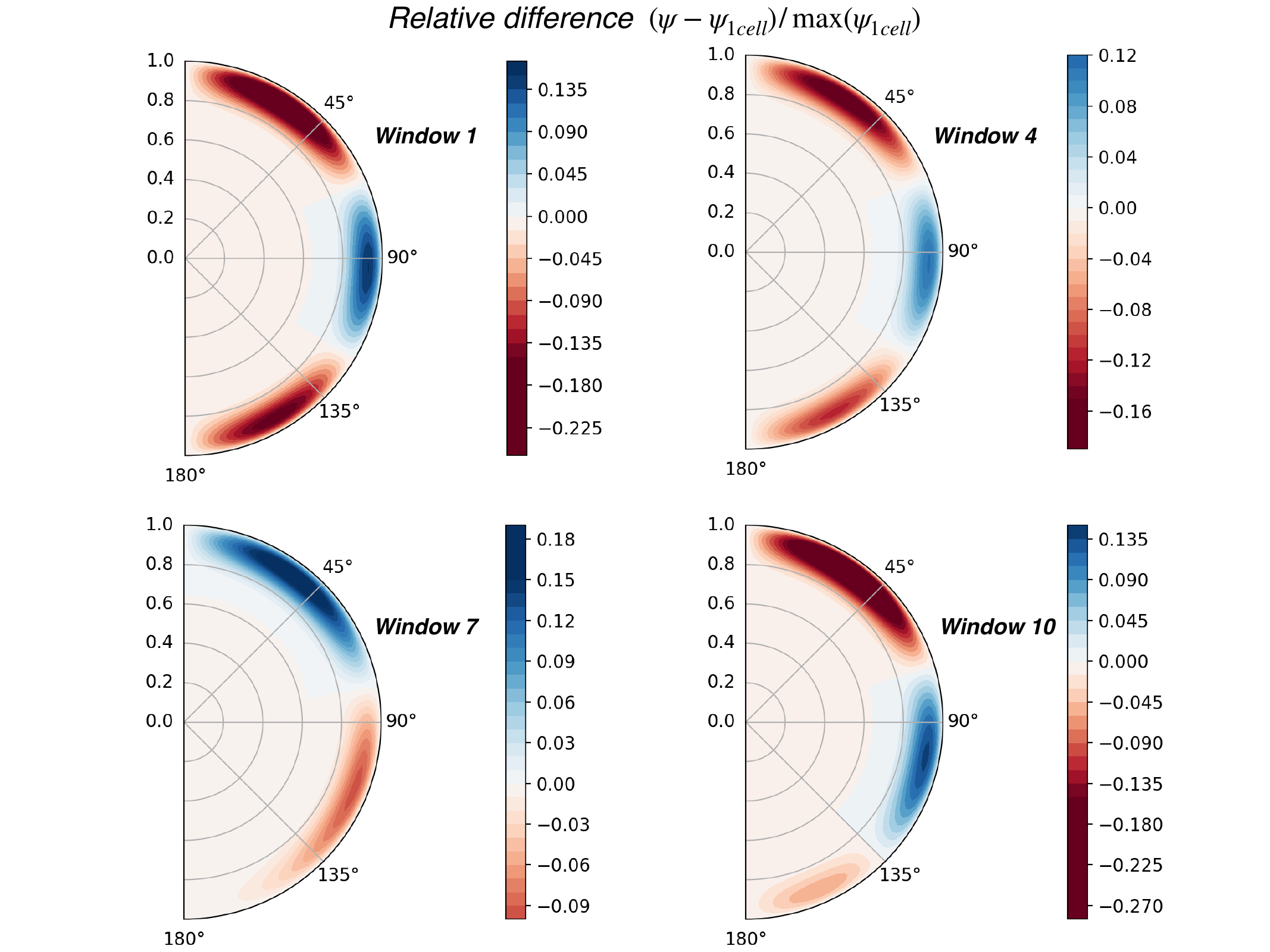}
 \caption{Difference in the meridional flow streamfunction between the result of the assimilation in one particular window and the basic 1-cell meridional flow. This is chosen from the assimilation phase of a case with $A_W=21$, representative of the mean (green curve shown in Figure \ref{Fig:cycle25hemi}). We see that the flow never departs by more than around $20\%$ from the 1-cell MC. }
\label{Fig:MC}
\end{figure}

\begin{figure}[h!]
\begin{center}
\includegraphics[width=0.25\textwidth]{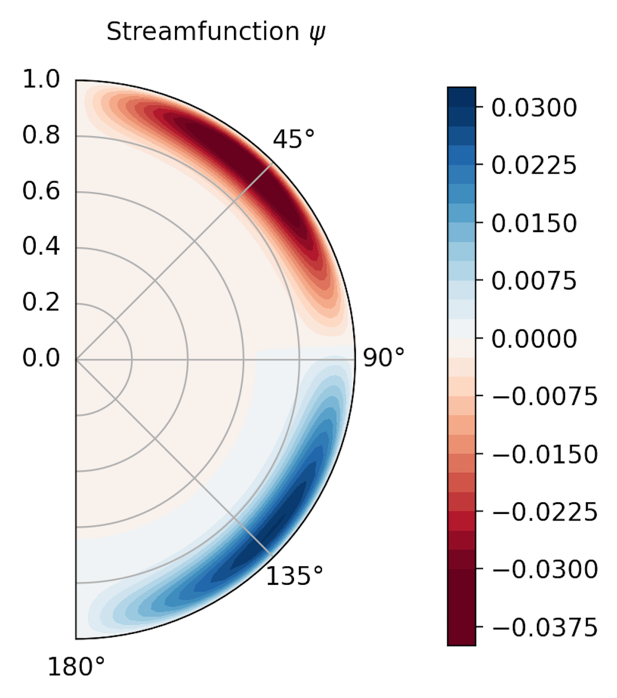}
\includegraphics[width=0.22\textwidth]{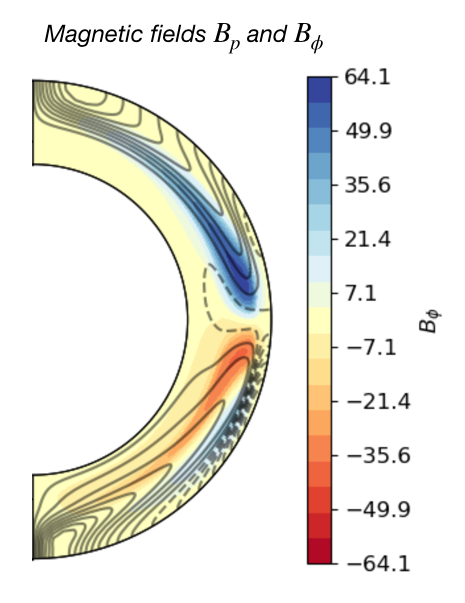}
\end{center}
 \caption{sMeridional flow streamfunction and magnetic initial conditions (toroidal field in colors and poloidal field lines in grey contours) obtained at the end of the last assimilation window (corresponding to window 10 of fig.\ref{Fig:MC}). This meridional circulation and initial magnetic conditions are then used to produce a forecast for cycle 25.}
\label{Fig:ab}
\end{figure}

We note that in April 2024, both hemispheres were forecasted to be slightly out of phase, with similar activity levels. During Summer 2024 the southern hemisphere started to desynchronize more than anticipated, with a sudden activity surge, so that a time lag of half a year is becoming more likely. To illustrate the level of asymmetry captured during the assimilation phase, we show in Figure \ref{Fig:MC} an example of the reconstructed flow streamfunction at 4 different windows during the assimilation phase. Note that the streamfunction of the last window (window 10) is the one used to produce the forecasted trajectory. Since the streamfunction does not depart by more than about $20\%$ of the 1-cell basic flow, we plotted the relative difference between our reconstructed streamfunction and this basic flow. This figure shows that we indeed get departures from a symmetric flow with respect to the equator, limited by the constraint we impose on the coefficient $d_{1,3}$ to be less than $10\%$ of $d_{1,2}$. Examining fig.\ref{Fig:ab} also enables to reveal some asymmetry in the MC streamfunction and in the magnetic field distribution. In this figure, the flow streamfunction and the initial magnetic toroidal field (colors) and poloidal field (contours) obtained by assimilating the last window (window 10) are shown in the meridian plane. As discussed before, this level of asymmetry in the flow and field is responsible for the different trajectories obtained for both hemispheres.

\subsection{Prediction of the next minimum}
\label{sec_min}

Since we are likely near or have reached the maximum of cycle 25, it is important to forecast the next phase of the current cycle. To this end, we present here our forecast for the next minimum of solar activity. Indeed, we have shown that our temporal horizon window is of the order of 10 years so we are well within this horizon for the next minimum.
We know that cycle 24 minimum was on December 2019, so simply adding 11 years to this date will 
provide a no cast prediction for the minimum of cycle 25 around December 2030/early 2031.
As shown in Figure \ref{Fig:cycle25tot}, we find that the minimum should be around Fall 2029. An estimate of the first and third quartiles and of the median and mean calculated on the 480 members of the ensemble give the distribution shown in appendix in Fig. \ref{fig:min}. From this distribution of predictions, we estimate our most probable next minimum to occur between early 2029 to Fall 2030, i.e. $2029.6_{-0.5}^{+1}$. We note that the error bars are not symmetric with respect to the mean value, making a minimum in 2030 most probable. This implies that cycle 25 would not be a long cycle, contrary to cycle 23 that lasted 12.4 years. We are currently forecasting that it will more likely last between 9.5 and 10.5 yr, i.e. closer to the length of cycles 21 and 22.
Recall that historical records show that the solar cycle duration varies between 9 and 13.7 years, so cycle 25 would be in the 50\% of cycles that are a bit shorter than the canonical 11 yr period.

The SSN value reached during that minimum is shown on the figure to be below 10, but it is likely not that significant given the relative large error bar. Moreover, we note that while during the assimilation phase the temporal shape of the model (blue triple dot curve) can be asymmetric, this is typically not the case for the forecast, that usually displays a rather sinusoidal temporal shape. We know from historical records that the rising vs declining phase of the solar cycle is not symmetric with respect to the maximum, being often much longer by several years. This empirical asymmetry aspect of the solar cycle, called the Waldmeier effect, is not fully taken into account in our current forecast (see model description in section \ref{sec_model}).

\subsection{Predictions of Cycle 25 as a function of the cycle phase}
\label{sec_cycle25phase}

\begin{figure}[h!]
\begin{center}
\includegraphics[width=0.48\textwidth]{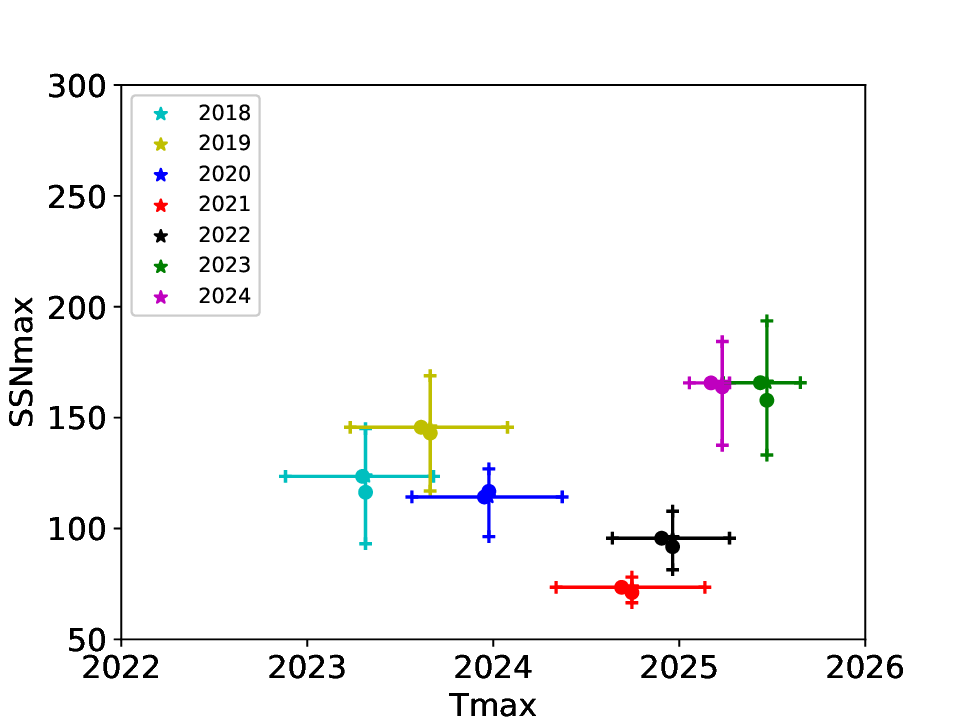}

\end{center}
\caption{Same as Figure \ref{fig_boxes} but for cycle 25. We here added 2 extra phases of the cycle, finishing in 2018 and 2019, i.e. before the previous minimum. }
\label{fig_box25}
\end{figure}

To further characterize our cycle 25 predictions, we now apply the convergence analysis as a function of the cycle phase, as it was quite informative to study it for cycles 22, 23 and 24.
The results are shown in Figure \ref{fig_box25} for an assimilation window extending up to 2018, 2019, 2020, 2021, 2022, 2023 and 2024. It has to be said here that because of the use of the 13-month smoothed data, the prediction shown here for 2024 actually uses data extending until July 2023. This explains why the predicted maximum here is slightly higher and later than the prediction presented in section \ref{sec:cyc25tot} where more updated data were used (observations until December 2023).

From this figure and comparing to the same plots for the previous cycles, it seems that cycle 25 predictions were always in favour of a weak cycle (SSN below 160) with small error bars for all cycle phases, showing quite accurate results in all cases. Incorporating more recent data (cycle phases 2023 and 2024) produces an increase in the prediction and a delay in time of the peak amplitude. It is also interesting to see that the timing of the maximum has shifted consistently when more and more updated data were incorporated in the \emph{Solar Predict Tool} and that the most recent predictions seem to converge on a maximum time at the end of 2024 / beginning of 2025.

Our results for the phase 2019 (yellow point on the figure with a mean max SSN of $145.7\pm 26$ reached in Mid 2023 with a six months uncertainty) may be compared to the prediction we published in \cite{Brun20} where a maximum SSN of $92 \pm 10$ was announced to be reached in April $2024 \pm 6$ months. At the time, we were using the 5-yr smoothed SSN time series and 10-yr smoothed butterfly diagram when we are here using the 13-month smoothed time series for the SSN and the 1-yr smoothed data for the butterfly diagram. We also extended here the ensemble by adding more values of $A_W$ and by introducing the regularization terms $J_1, J_2$ and $J_3$. Those various changes contribute to a different prediction now reached starting from 2019 compared to \cite{Brun20}. This shows again the importance of carefully choosing the assimilated data and the various terms in the objective function for a reliable prediction.

\section{Discussion and conclusion}
\label{sec_conclu}

In this work, we presented a first attempt to apply our variational data assimilation framework on real solar data provided by the SIDC SILSO database and synoptic maps coming from various solar observatories. The assimilated data consists in the sunspot number time series smoothed over 13 months and the surface line-of-sight magnetic field coming from the synoptic maps over one full cycle. The output of our assimilation framework is the prediction of cycle 25 length and maximum values, by adjusting a dynamo model to the observed data. 
One strength of the method presented here is that an ensemble of predictions are produced: a first ensemble is produced by varying the different parameters of the assimilation, namely the weighting coefficients in front of the terms appearing in the objective function to be minimized. A second ensemble is then produced by perturbing the results of our assimilation, namely a best-fit meridional circulation and initial magnetic conditions for the forecast. This ensemble of hundreds of predictions then enables us to precisely calculate the mean, median and various quantile values of the distribution of maximum sunspot number and timing of this max and to estimate error bars on our predictions. Finally, an originality of this work is to provide separate predictions for the Northern and Southern hemispheres, since an asymmetry parameter allows our dynamo model to produce magnetic fields with different behaviours in both hemispheres.

 We tested our method on previous cycles for the sake of validation and find reasonable agreements between predicted and observed maximum values and timing for cycles 22, 23 and 24. The results for the prediction of cycle 22 are however less satisfactory, possibly owing to the fact that the previous minimum extended for almost 2 years between 1985 and 1987. Using data smoothed on longer time-scales could be a way to improve the forecasts in those particular cases. It has to be noted that most cycles present two peaks owing to the fact that the two hemispheres reach their maximum at different dates sometimes separated by several months or even years. Sometimes, the two peaks can reach similar amplitudes. We can then wonder which value would indeed correspond to the true maximum and if the prediction could be considered as successful if the forecast would peak between the two real values. This also questions the possibility of an axisymmetric mean-field dynamo model to reproduce a high degree of asymmetry between the two hemispheres. We indeed find in this work that our forecasted trajectories currently fail at reproducing large asymmetries between the hemispheres. Possible ways forward would be to enable more coefficients in the meridional flow expansion to be assimilated and to relax the strong constraint we imposed of the $d_{1,3}$ coefficient (the main driver of North/South asymmetries) to be less than $10\%$ of the 1-cell coefficient $d_{1,2}$. Progress may be needed in this direction in the future.  
 
 As far as the ongoing cycle 25 is concerned, our prediction for the maximum, as of writing, gives a sunspot maximum of $143.1 \pm 15.0$ reached at the end of 2024, with similar amplitudes between the Northern and Southern hemispheres. Our method shows predictions varying with the extent of the assimilation window, but becoming more precise and accurate as more and more updated data are assimilated into our model. However, the predicted values may still depend on the level of filtering of the observed SSN and butterfly diagram and on our various parameters. We here focused on the most significant influence of the weight put on the SSN observations compared to the surface radial field but additional investigation remains to be conducted for other parameters which may influence the quality of these predictions.s

Given that we have reached the maximum of cycle 25 at the time of writing of this study (Fall 2024), it is instructive to also forecast the next minimum. This is a reasonable goal since the temporal horizon window of our pipeline is around 10 years (as shown in \cite{Hungetal2017ApJ}), so far enough to attempt such forecast which is likely occurring within the next 7 years or so. We have shown that our tool is forecasting the next minimum to occur in late 2029 with a 1-yr error bar. A no-cast forecast just adding 11 years to the last minimum that occurred in December 2019, would imply a minimum in late 2030/early 2031, so significantly later. Hence, our tool indicates that cycle 25 would be a relatively short cycle, with a length ranging from 9.5 to 10.5 years, that puts it in the lower 50\% of the distribution of all previous cycle lengths on records. Given that our forecast is not fully taking into account the empirical Waldmeier rule we will need to work more specifically on better incorporating this asymmetry between rising and declining phases of the solar cycle, as in \cite{2012A&A...542A..26P} for instance, to have more robust forecasting overall, improvement we intend to work on in the near future.

 In addition, several aspects of our method may now be improved, in particular for the forecasting part. First we could consider other sources of poloidal field like an $\alpha$-effect distributed across the whole convection zone, which would make the link between the simulated toroidal field at the base of the convection zone and the observed sunspot number less direct. We could also implement additional transport mechanisms like turbulent pumping, which would reduce the dependency of the cycle period to the meridional flow speed as shown in \cite{DoCao11} or \cite{Karak12} for example. Another major improvement would be to introduce a back-reaction of the Lorentz force on the fluctuations of the velocity field, in the spirit of the Malkus-Proctor effect or $\Lambda$-effect \citep{Malkus75, Kichatinov93, Bushby06} to better take into account non-linearities which are still weak in our current model. These additional non-linearities would naturally lead to modulations of the cycle period and amplitude and may be important to take into account for the predictive part of our technique. Another way forward is related to our modelling of emergence of magnetic flux. Indeed, as flux emergence at the solar surface may also play a crucial role in the timing and value of the maximum of solar cycles \citep[e.g.][]{CameronSchuessler07}, additional observational constraints on the tilt angles, amplitudes of the sunspot magnetic fields and possibly morphological characteristics of sunspots could be introduced in the objective function to improve our assimilation step and consequently the accuracy of our forecast of future solar activity. This would also imply to use a direct dynamo model which better takes into the account the individual characteristics of active regions instead of their average effect on the cycle, as the double-ring approach first suggested by \cite{Durney95} and then implemented for example in \cite{Nandy01} or even a 3D kinematic version of the Babcock-Leighton model as used in \cite{Yeates13} or \cite{Kumar19}. These additional developments of the dynamo model should definitely be considered in future data assimilation techniques applied to solar physics to improve our forecasting capabilities.

 Our \emph{Solar Predict tool} is freely available on the ESA SWE portal under the following link: \href{https://swe.ssa.esa.int/irfu-federated}{https://swe.ssa.esa.int/irfu-federated}.

\begin{acknowledgements}
We acknowledge funding support by ERC grants Whole Sun \#810218 and Solar Predict \#640997, ESA SWESNET support, INSU/PNST grant and CNES Space Weather financial supports. We are thankful to WDC-SILSO, Royal Observatory of Belgium, Brussels, for providing free access to their sunspot number times series used in our data assimilation pipeline. Likewise, this work utilizes GONG data obtained by the NSO Integrated Synoptic Program, managed by the National Solar Observatory, which is operated by the Association of Universities for Research in Astronomy (AURA), Inc. under a cooperative agreement with the National Science Foundation and with contribution from the National Oceanic and Atmospheric Administration. The GONG network of instruments is hosted by the Big Bear Solar Observatory, High Altitude Observatory, Learmonth Solar Observatory, Udaipur Solar Observatory, Instituto de Astrofísica de Canarias, and Cerro Tololo Interamerican Observatory. We also use synoptic maps from the Wilcox Observatory for data earlier than 2000 and in case of gaps we also use SOLIS data. SOLIS data are obtained by the NSO Integrated Synoptic Program, managed by the National Solar Observatory, which is operated by the Association of Universities for Research in Astronomy (AURA), Inc. under a cooperative agreement with the National Science Foundation. AS acknowledges funding from the French Agence Nationale de la Recherche (ANR) project STORMGENESIS \# ANR-22-CE31-0013-01. We are grateful to Eric Buchlin and Stephane Caminade for their help in deploying and hosting the \emph{Solar Predict tool} on the IAS/Medoc servers \href{https://idoc.osups.universite-paris-saclay.fr/medoc/}{https://idoc.osups.universite-paris-saclay.fr/medoc/}.
\end{acknowledgements}

\bibliographystyle{./aa}
\bibliography{mybib}

\appendix

\section{Extension of hemispherical SSN times series}

\begin{figure}[h!]
\begin{center}
\includegraphics[width=0.48\textwidth]{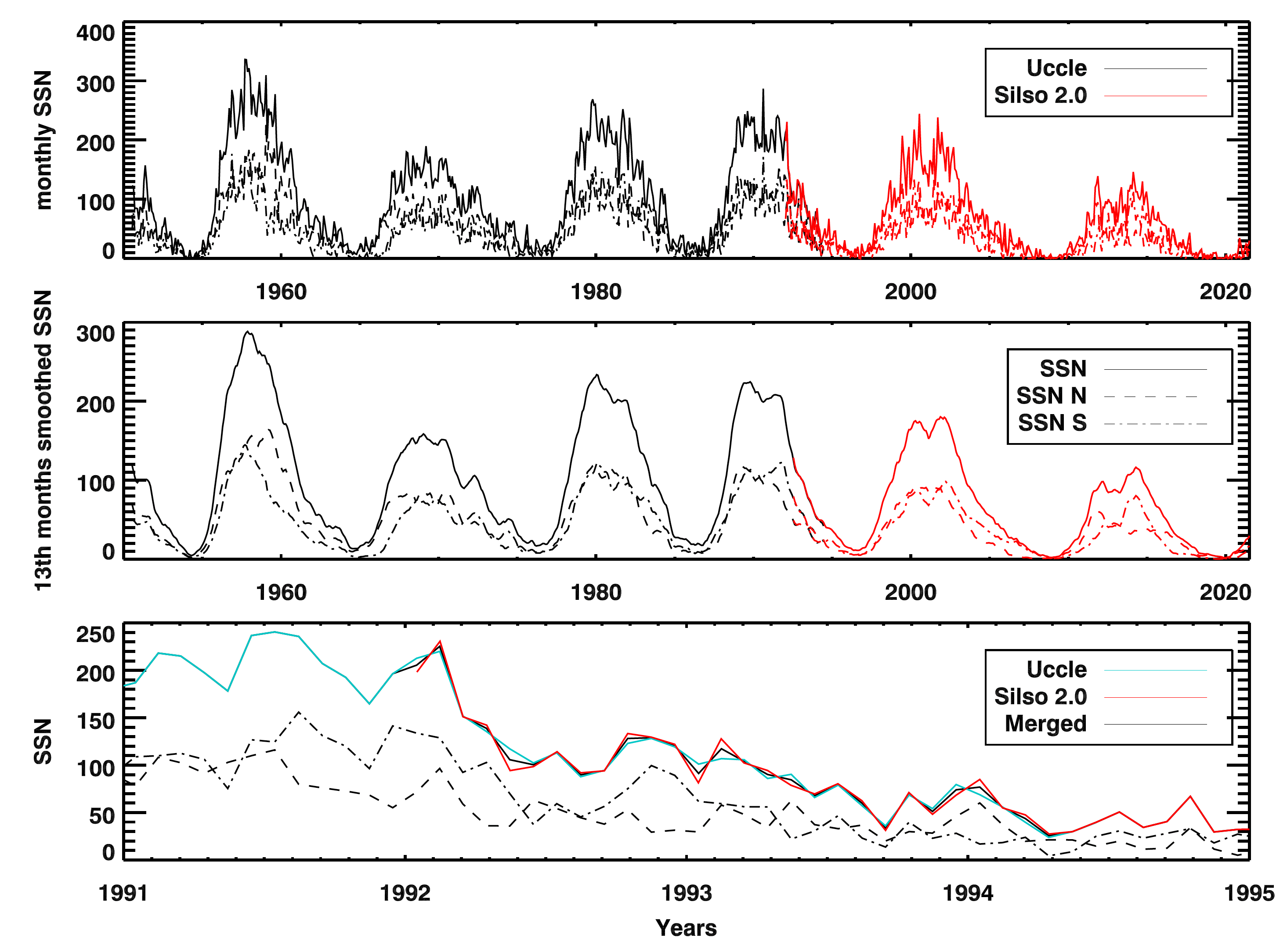}

\end{center}
\caption{SILSO SSN hemispherical time series extension using UCCLE station at ROB. On the top panel we shown the monthly SSN time series (using black color for UCCLE and red for SILSO), in the middle panel the 13-month smoothed equivalent SSN time series and in the bottom panel the merging of the time series around 1992 using cyan color for UCCLE, red for SILSO and black color for the merged time series.}
\label{fig:ucle}
\end{figure}

In order to extend in the past (i.e. 1975) the hemispherical time series of SILSO (that starts in 1992) we have made use of the UCCLE station SSN time series. The UCCLE station is run by ROB and in the file we got runs from 1950 until 1994, hence overlapping for 2 years with the SILSO hemispherical time series. We have found that a correction factor around 1.4136 was needed to cross-calibrate the data (i.e. the UCCLE data was boosted by this factor). We show both times series merged in Figure \ref{fig:ucle}, with on the top panel the monthly SSN time series, in the middle panel the corresponding 13-month smoothed SSN time series and in the bottom panel the merging of both 13-month smoothed time series around year 1992.

\section{Prediction for the next minium}

\begin{figure}[h!]
\begin{center}
\includegraphics[width=0.48\textwidth]{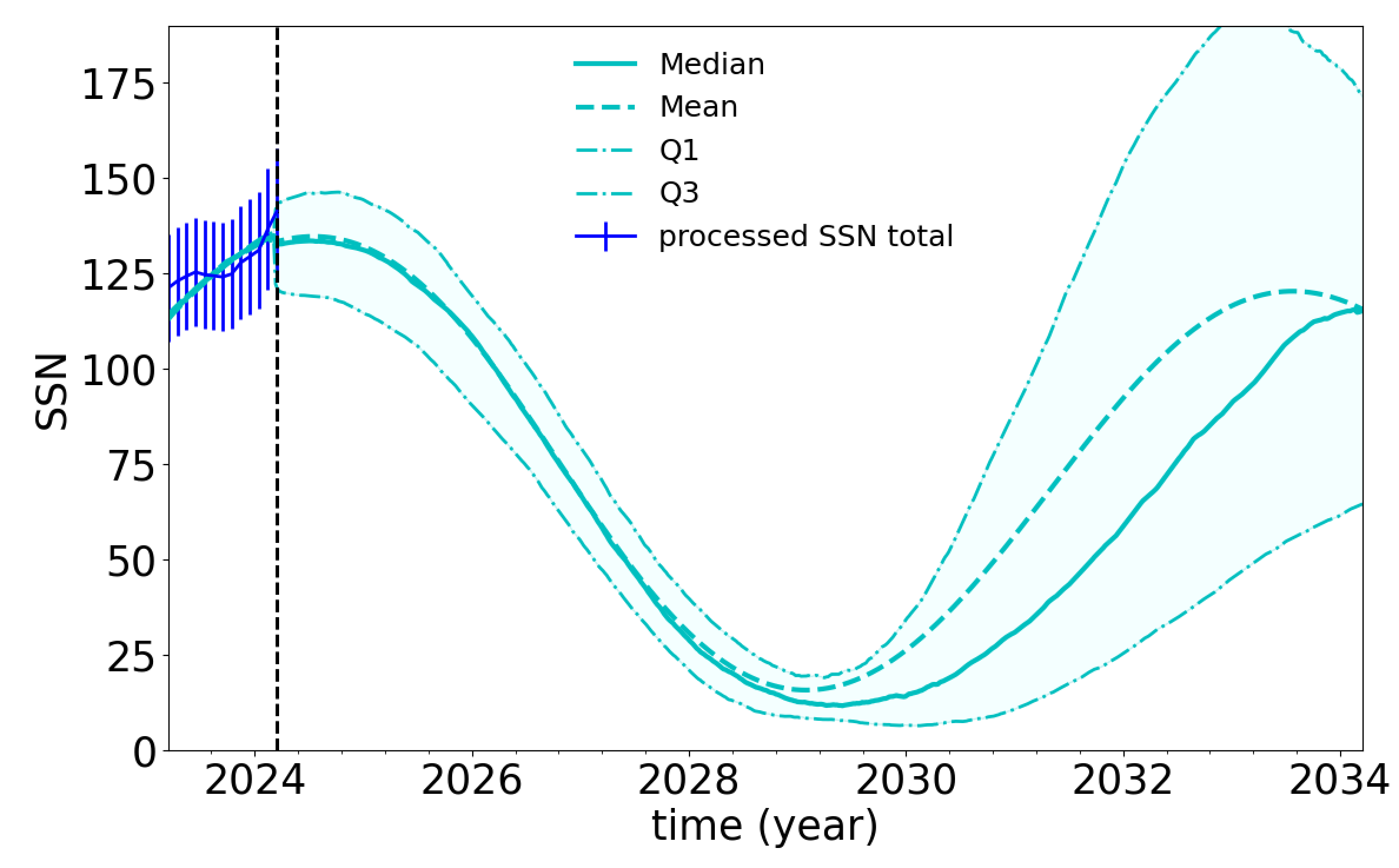}
\end{center}
\caption{sPrediction of the next solar minimum, with the mean, median and first and third quartiles of the distribution.}
\label{fig:min}
\end{figure}

To better estimate the typical error in the prediction of the next solar minimum discussed in Sec. \ref{sec_min}, we calculated the distribution of predictions made using the 480 members described in the main text, when assimilation is performed up to April 2024. Figure \ref{fig:min} shows the mean, median and first and third quartiles of the distribution. The calculation of these values allows us to have a more precise estimate of the next minimum predicted to occur in $2029.6_{-0.5}^{+1}$, however with an error bar asymmetric with respect to the mean value, indicating in fact a probable minimum in 2030.

\section{Summary of cycles 22, 23, 24 and 25 forecasts}

In this section we summarize the values of the forecasts for cycles 22, 23, 24 and 25. Table \ref{tab} gives the values of the forecasted value and timing of the maximum SSN for different assimilation windows for the past 3 cycles and the current one. In bold characters we indicate the mean values and the 3 quartiles are also given.

\begin{table*}
\centering
\begin{tabular}{c c c c c c c c c}
\hline 
\hline  
Year at end of assimilation& {\bf Mean($t_{max}$)} & $q1_{tmax}$ & $q2_{tmax}$   & $q3_{tmax}$ & {\bf Mean($\rm SSN_{max}$)} & $q1_{ssnmax}$ & $q2_{ssnmax}$ & $q3_{ssnmax}$ \\ 
\hline \\
1985 & {\bf 1990.537 } & 1989.868& 1990.456 & 1991.398 &  {\bf 117.273} & 75.800 & 99.413 & 127.321\\ 
\hline
1986 & {\bf 1992.214} & 1991.799& 1992.264 & 1992.612 &  {\bf 159.652} & 123.208 & 152.666 & 184.150\\ 
\hline
1987 & {\bf 1991.432 }& 1989.831& 1990.705 & 1991.352 &  {\bf 147.391} & 88.600 &115.132 & 202.138\\ 
\hline
1988 & {\bf 1990.955} & 1989.545& 1990.705 & 1991.352 &  {\bf 196.550} & 116.077 & 181.352 & 266.666\\ 
\hline
1989 & {\bf 1990.232} & 1989.677& 1990.307 & 1990.556 &  {\bf 251.761} & 230.093 & 242.797 & 270.112\\ 
\hline
Observed cycle 22 & {\bf 1989.54} & & & &  {\bf 225.36 $\pm$ 15.9} &  &  & \\ 
\hline
\hline
\hline \\
1996 & {\bf 2001.555 } & 2001.224& 2001.796 & 2001.978 &  {\bf 149.352} & 83.631 & 100.979 & 257.977\\ 
\hline
1997 & {\bf 2001.818} & 2001.464& 2001.821 & 2002.177 &  {\bf 209.076} & 166.785 & 222.288 & 265.186\\ 
\hline
1998 & {\bf 2001.632 }& 2001.166& 2001.663 & 2002.094 &  {\bf 200.924} & 123.404 &210.913 & 269.167\\ 
\hline
1999 & {\bf 2001.525} & 2001.398& 2001.514 & 2001.663 &  {\bf 204.667} & 168.043 & 202.682 & 233.029\\ 
\hline
2000 & {\bf 2001.321} & 2001.249& 2001.331 & 2001.398 &  {\bf 190.189} & 159.884 & 189.659 & 213.428\\ 
\hline
Observed cycle 23 & {\bf 2001.784} & & & &  {\bf 180.3 $\pm$ 10.8} &  &  & \\ 
\hline
\hline
\hline\\
2007 & {\bf 2012.631} & 2012.136& 2012.580 & 2013.085 &  {\bf 143.292} & 65.701 & 135.848 & 216.565\\ 
\hline
2008 & {\bf 2013.287} & 2012.886& 2013.284 & 2013.698 &  {\bf 113.697} & 90.700 & 110.213 & 129.576\\ 
\hline
2009 & {\bf 2013.673} & 2013.317& 2013.665 & 2014.014 &  {\bf 87.607} & 75.500 & 86.430 & 95.671\\ 
\hline
2010 & {\bf 2013.895} & 2013.574& 2013.881 & 2014.179 &  {\bf 87.266} & 73.709 & 84.682 & 98.804\\ 
\hline
2011 & {\bf 2013.987} & 2013.699& 2013.939 & 2014.96 &  {\bf 108.873} & 87.911 & 109.100 & 128.893\\ 
\hline
Observed cycle 24 & {\bf 2014.288} & & & &  {\bf 116.4 $\pm$ 8.2} &  &  & \\ 
\hline
\hline
\hline\\
2018 & {\bf 2023.313} & 2022.883& 2023.297 & 2023.679 &  {\bf 123.508} & 93.137 & 116.347 & 145.028\\ 
\hline
2019 & {\bf 2023.661} & 2023.231& 2023.952 & 2024.076 &  {\bf 145.653} & 116.907 & 143.030 & 168.903\\ 
\hline
2020 & {\bf 2023.976} & 2023.563& 2023.952 & 2024.370 &  {\bf 114.213} & 96.351 & 116.773 & 126.916\\ 
\hline
2021 & {\bf 2024.745} & 2024.337& 2024.690 & 2025.137 &  {\bf 73.484} & 66.546 & 71.138 & 78.088\\ 
\hline
2022 & {\bf 2024.964} & 2024.640& 2024.905 & 2025.270 &  {\bf 95.622} & 81.412 & 91.809 & 107.788\\ 
\hline
2023 & {\bf 2025.471} & 2025.237& 2025.436 & 2025.651 &  {\bf 165.744} & 133.140 & 157.813 & 193.623\\ 
\hline
2024 & {\bf 2025.231} & 2025.054& 2025.170 & 2025.270 &  {\bf 165.655} & 137.522 & 163.838 & 184.307\\ 
\hline
\hline
\end{tabular}
\caption{Results for the prediction of cycles 22, 23, 24 and 25. The values $q1, q2$ and $q3$ indicate respectively the 1st, 2nd and 3rd quartile of the distributions of $t_{max}$ and $\rm SSN_{max}$. We also indicate the observed values for both quantities for cycles 22 to 24.} %From the 13-month filtered SSN monitoring of SIDC, real values for Cycle 22 is $\bf 1989.54$ and  $\bf 225.36 \pm 15.9$, for Cycle 23 $\bf 2001.784$ and $\bf 180.3 \pm 10.8$ and for Cycle 24 $\bf 2014.288$ and $\bf 116.4 \pm 8.2$.} 
\label{tab} % is used to refer this table in the text
\end{table*}

\end{document}